\documentclass[journal]{IEEEtran}

\usepackage[utf8]{inputenc}
\usepackage{amsmath}
\usepackage{amsfonts}
\usepackage{amssymb}
\usepackage{graphicx}
\usepackage{subcaption}
\usepackage{authblk}
\usepackage{color}
\usepackage{url}

\newcommand{\e}{\text{error}}

\newcommand{\I}{\iota}
\newcommand{\rev}[1]{{\color{black}{#1}}}
\newcommand{\revv}[1]{{\color{black}{#1}}}

\numberwithin{equation}{section}

\begin{document}

\title{Unrolled algorithms for group synchronization}

\author[1]{Noam Janco}
\author[1]{Tamir Bendory\thanks{This work was partially supported by the  NSF-BSF grant 2019752, the BSF grant no. 2020159, and the ISF grant no. 1924/21. }}

\affil[1]{School of Electrical Engineering,	Tel Aviv University, Tel Aviv, Israel}


\maketitle

\begin{abstract}
The group synchronization problem involves
estimating a collection of 
group elements from noisy measurements of their pairwise ratios. 
This task is a key component in many computational problems, including the molecular reconstruction problem in single-particle cryo-electron microscopy (cryo-EM).
The standard methods to estimate the group elements  are based on iteratively applying linear and non-linear operators, \revv{and are not necessarily optimal.} 
Motivated by the structural similarity to deep neural networks, we adopt the concept of algorithm unrolling,
where training data is used to optimize the algorithm. 
We design unrolled algorithms for several group synchronization instances, including synchronization over the group of 3-D rotations: the synchronization problem in cryo-EM. We also apply a similar approach to the multi-reference alignment problem. We show by numerical experiments that the unrolling strategy outperforms existing synchronization algorithms in a wide variety of scenarios. 

\end{abstract}

\begin{IEEEkeywords}
Group synchronization, algorithm unrolling, multi-reference alignment
\end{IEEEkeywords}

\section{Introduction}

\IEEEPARstart{G}{iven} a  group $G$, the group synchronization problem entails estimating $N$ elements $g_1,\ldots, g_N\in G$ from their noisy pairwise ratios $g_{ij}\approx g_ig_j^{-1}$. Since $g_ig_j^{-1}  = (g_ig)(g_jg)^{-1}$ for any $g\in G$, the group elements can be estimated up  to a right multiplication by some $g\in G$.
A canonical example is the angular synchronization problem of estimating 
\(N\)  angles \(\theta_1,\dots,\theta_N\in\left[0,2\pi\right)\) from   their noisy  offsets $\theta_{ij}\approx(\theta_i-\theta_j) \bmod 2\pi$; this problem corresponds to  synchronization over  the group of complex numbers on the unit circle $U(1)$~\cite{singer2009angular,boumal2016nonconvex,bandeira2017tightness,zhong2018near}.

Under the standard  additive Gaussian noise model,  
the maximum likelihood estimator (MLE) of the angular synchronization problem can be formulated as the solution of a non-convex optimization problem on the manifold of product of circles: 
\begin{equation}
\label{cost_ang_synchronization}
\max_{z\in \mathbb{C}_1^N}{z^*Hz},
\end{equation}
where \(H_{ij}=e^{\I\theta_{ij}}\) is the measurement matrix, $\I=\sqrt{-1}$, and \( \mathbb{C}_1^N := \{z\in\mathbb{C}^N:|z_1|=...=|z_N|=1\}\).
Singer~\cite{singer2009angular} proposed to solve~\eqref{cost_ang_synchronization} by extracting the leading eigenvector of $H$ using the power method:  
given an initial estimate of the sought angles, the power method  iteratively applies the matrix $H$ to the current estimate,  and then normalizes its norm.
In follow-up papers, Boumal~\cite{boumal2016nonconvex} suggested an alternative normalization strategy, and Perry et al.~\cite{perry2018message} developed an algorithm which is inspired by the approximate message passing (AMP) framework.
These strategies can be naturally extended to additional group synchronization setups. 
We describe all these methods in detail in Section~\ref{sec:existing_methods}. 
For our purposes, it is important to note that  the $t$-th iteration of all these methods follow the same  structure: 
\begin{equation}
    \label{general_iteration}
    z^{(t)} = f(H,z^{(t-1)}, z^{(t-2)}),
\end{equation}
for some non-linear function $f$.
Specifically, at each iteration, the current estimate is acted upon by a linear operator, followed by a non-linear function. This structural resemblance to the blueprint of a neural network layer is the cornerstone of this work. 

The group synchronization problem is an important component in a variety of scientific, engineering, and mathematical problems, including the structure from motion problem~\cite{ozyecsil2017survey},
sensor  network localization~\cite{cucuringu2012sensor}, phase retrieval~\cite{marchesini2016alternating,iwen2016fast,bendory2017non}, ranking~\cite{cucuringu2016sync}, community detection~\cite{abbe2017community},
and 
 synchronization of the  rigid motion group~\cite{rosen2019se,ozyesil2018synchronization,briales2017cartan,bendory2021compactification}, the dihedral group~\cite{bendory2022dihedral},   and the permutation group~\cite{ling2022near}.
In Section~\ref{sec:conclusion} we discuss how the proposed algorithm for synchronization over the group of 3-D rotations can be applied to the molecular reconstruction problem in single-particle
cryo-electron microscopy (cryo-EM)~\cite{singer2011three,shkolnisky2012viewing,bendory2020single}.

Motivated by the fact that  existing synchronization methods are not optimal, and the resemblance of the iteration~\eqref{general_iteration} to the general structure of a  modern neural network layer, we adopt the approach of algorithm unrolling~\cite{monga2021algorithm}, to develop an efficient, interpretable neural network that outperforms  existing methods.
The underlying idea of  algorithm unrolling, first introduced in the seminal work of Gregor and LeCun~\cite{gregor2010learning},  is to exploit existing iterative algorithms and optimize them using training data. 
Specifically, each iteration of the algorithm  is represented as a layer of a  network, and concatenating these layers forms a deep neural network. Passing through the network is analogous to executing the iterative algorithm for a fixed number of steps. The network can be trained using back-propagation, resulting in model parameters that are learned from training samples. Thus, the trained network can be naturally interpreted as an optimized algorithm. 
This is especially important since, while the past decade has witnessed the unprecedented success of deep learning techniques in numerous applications, most  deep learning techniques are purely data-driven, and the underlying structures are hard to interpret. 
The unrolled networks are parameter efficient, require less training data, and less susceptible to overfitting. 
Moreover, the unrolled networks naturally inherit prior structures and domain knowledge, leading to better generalization.
The algorithm unrolling  approach has been adopted to various tasks in recent years,  including compressive sensing~\cite{yang2018admm}, image processing~\rev{\cite{li2019algorithm,chen2016trainable,Wu2019-gk,9152164,9123435}}, graph signal processing~\cite{chen2021graph}, biological imaging~\cite{sahel2022deep}, to name but a few.
We refer the readers to a recent survey on algorithm unrolling and references therein~\cite{monga2021algorithm}.
Figure~\ref{z_2_unrolling_concept} demonstrates the concept of algorithm unrolling for the  synchronization problem over the group $\mathbb{Z}/2$; see Section~\ref{sec:existing_methods}.

We also study the application of the unrolling approach to the multi-reference alignment (MRA) problem. 
 MRA  is the problem  of estimating a signal from its multiple noisy copies, each  acted upon by a random group element. 
 The computational and statistical properties of the MRA problem have been analyzed thoroughly in recent years; see~\cite{b2013multireference,BispectrumInversion,bandeira2020optimal,perry2019sample,abbe2018multireference,ma2019heterogeneous,boumal2017heterogeneous, bandeira2020non, bandeira2017estimation, romanov2021multi,abas2022generalized,janco2022accelerated}.
\revv{Group synchronization is often used to solve the MRA problem in the high SNR regimes, by first estimating the pairwise ratios between the group elements from the noisy observations, and then estimating the group elements themselves as a synchronization problem.}
Given an accurate estimate of the random group elements,  the MRA problem reduces to a linear inverse problem, which is much easier to solve. 
 Importantly, in contrast to group synchronization, the goal in MRA is to estimate the underlying signal, while the group elements are nuisance variables whose estimation is merely an intermediate step.  

The rest of the paper is organized as follows. In Section~\ref{sec:existing_methods} we introduce three particular cases of group synchronization and two  MRA models, and present existing methods to solve them. 
Section~\ref{sec:unrolling} introduces the proposed unrolled algorithms, and Section~\ref{sec:results} shows numerical results. Finally, Section~\ref{sec:conclusion} concludes the paper, and outlines how the proposed synchronization technique over SO(3) can be applied to the reconstruction problem in cryo-EM.

\section{Group synchronization, multi-reference alignment, and existing solutions}
\label{sec:existing_methods}
In this section, we introduce three  group synchronization and two MRA  models. 
We also elaborate on three different methods to estimate the group elements. 
These methods are the keystone of the unrolled algorithms described in the next section.

\subsection{$\mathbb{Z}/2$ synchronization}
\label{z_2_existing_methods}
We begin with the simplest group synchronization problem over the group $\mathbb{Z}/2$. The goal is to estimate a  signal \(z\in\{\pm1\}^N\) from the noisy measurement matrix:
\begin{equation}
\label{z_2_gaussian_model}
H = \frac{\lambda}{N}zz^T + \frac{1}{\sqrt{N}}W,
\end{equation}
where \rev{$W_{ij}=W_{ji}\sim \mathcal{N}(0,1)$}, and \(\lambda>0\) is a signal-to-noise ratio (SNR) parameter. 
\rev{The scaling is such that the signal and noise components of the observed data are of
comparable magnitudes.}
The diagonal entries of $W$  follow the same distribution.
We also assume that each entry of $z$ is drawn i.i.d.\ from a uniform distribution over $\pm 1$. 
We can only hope to estimate \(z\) up to a sign, due to the symmetry of the problem. 

The $\mathbb{Z}/2$ synchronization problem is associated with the maximum likelihood estimation problem:
\begin{equation}
\label{cost_synchronization_z_2}
\max_{z\in \mathbb{R}_1^N}{z^THz},
\end{equation}
where \( \mathbb{R}_1^N := \{z\in\mathbb{R}^N:|z_1|=...=|z_N|=1\}\). This is a non-convex optimization problem. We now describe different existing iterative algorithms to solve~\eqref{cost_synchronization_z_2}. All algorithms are initialized  with small random values in $[-1,1]$. 
Specifically, in our numerical experiments, the algorithms are initialized by  $z^{(0)},z^{(-1)}\sim \mathcal{N}(0,10^{-2}I)$.

\subsubsection{Power method (PM)}
In~\cite{singer2009angular}, Singer proposed a spectral  approach (in the context of $U(1)$ synchronization) that relaxes~\eqref{cost_synchronization_z_2}  to
\begin{equation} \label{eq:Rayleigh}
\max_{z\in\mathbb{R}^N, \|z\|^2 = N}{z^THz} = \max_{z\in\mathbb{R}^N, \|z\|^2 = N}{N\frac{z^THz}{\|z\|^2}}.
\end{equation}
The expression in~\eqref{eq:Rayleigh} is known as  the Rayleigh quotient and is maximized by  the leading eigenvector of \(H\) that corresponds to the largest eigenvalue. This eigenvector can be computed using the power  method, whose $(t+1)$-th iteration reads:
\begin{equation}
\label{power_iteration_z_2}
z^{(t+1)}=\frac{Hz^{(t)}}{\|Hz^{(t)}\|}.
\end{equation}
After the last iteration $T$, the output is projected onto the $\mathbb{Z}/2$ group by $z(T)=\text{sign}(z(T))$, where \(\text{sign}()\) is the sign function,  acting separately on each entry of the vector.

\subsubsection{Projected power method (PPM)} \label{sec:sync_z2_ppm}
The projected power method~\cite{boumal2016nonconvex} suggests to replace the global normalization~\eqref{power_iteration_z_2} by an entrywise projection onto the group. 
Specifically, the $(t+1)$-th iteration reads:
\begin{equation} \label{eq:ppm_z2}
z^{(t+1)}=\text{sign}(Hz^{(t)}).
\end{equation}

\subsubsection{Approximate message passing (AMP)}
Perry et al.~\cite{perry2018message} proposed an algorithm which is inspired by the AMP framework. 
For the $\mathbb{Z}/2$ synchronization, its  $(t+1)$-th iteration reads:
\begin{equation}
\label{z_2_amp_1}
	z^{(t+1)}=\tanh(c^{(t+1)}),
\end{equation}
where
\begin{equation}
\label{z_2_amp_2}
c^{(t+1)}=\lambda Hz^{(t)} - \lambda^2(1-\langle(z^{(t)})^2\rangle)z^{(t-1)},
\end{equation}
and \(\langle\cdot\rangle\) denotes averaging over the vector entries.
The second term in~\eqref{z_2_amp_2} is called the Onsager correction term and is related to  backtracking messages in the graphical model~\cite{perry2018message}.

We underscore that all the methods mentioned above share a similar structure: the current  estimate of the group elements is multiplied by the measurement matrix, followed by a non-linear function.

\subsection{$U(1)$ synchronization}
\label{u_1_existing_methods}
Next, we consider the synchronization problem  over the  group  \(U(1)\) of complex numbers with unit modulus. 
The goal is to estimate $N$ elements \(z\in \mathbb{C}_1^N\), given the measurement matrix
\begin{equation}
\label{u_1_gaussian_model}
H = \frac{\lambda}{N}zz^* + \frac{1}{\sqrt{N}}W,
\end{equation}
where \(W\) is a Hermitian matrix whose entries are distributed independently (up to symmetry) according to the standard complex normal distribution \(\mathcal{CN}(0,1)\), and \(\lambda>0\) is an SNR parameter.
The diagonal entries of $W$ are drawn from the same distribution.
We assume that each entry of $z$ is drawn i.i.d.\ from a uniform distribution on the unit circle.
Due to symmetry considerations, we can only hope to estimate \(z\) up to a global element of $U(1)$. 
This synchronization problem is associated with the maximum likelihood estimation problem:
\begin{equation}
\label{cost_synchronization}
\max_{z\in \mathbb{C}_1^N}{z^*Hz}.
\end{equation}
This is a smooth, non-convex optimization problem on the manifold of product of circles.
We describe different existing iterative algorithms to maximize~\eqref{cost_synchronization}.  All algorithms are initialized  with small random values on the unit circle. In our experiments, the algorithms are initialized with  $z^{(0)},z^{(-1)}\sim \mathcal{CN}(0,2\cdot10^{-4}I)$.

\subsubsection{Power method (PM)}
Using a relaxation similar to~\eqref{eq:Rayleigh} with $z\in\mathbb{C}^N$ instead of $z\in\mathbb{R}^N$, we get power iterations as in \eqref{power_iteration_z_2}.

\subsubsection{Projected power method (PPM)}
Similarly to~\eqref{eq:ppm_z2},  the \((t+1)\)-th iteration of the PPM reads:
\begin{equation}
z^{(t+1)}=\text{phase}(Hz^{(t)}),
\end{equation}
where \(\text{phase}(z)[i]=z[i]/|z[i]|\).

\subsubsection{Approximate message passing (AMP)}
Following~\cite{perry2018message}, for each $i=1, \ldots,N$, the $(t+1)$-th iteration of the AMP algorithm reads:
\begin{equation}
	\label{u_1_amp_2}
	z^{(t+1)}[i]=f\left(|c^{(t+1)}[i]|\right) \frac{c^{(t+1)}[i]}{|c^{(t+1)}[i]|}, 
\end{equation}
where \(f(t)=I_1(2t)/I_0(2t)\), \(I_k\) denotes the modified Bessel functions of the first kind of order $k$, and
\begin{equation}
\label{u_1_amp_1}
c^{(t+1)}=\lambda Hz^{(t)} - \lambda^2(1-\langle|z^{(t)}|^2\rangle)z^{(t-1)}.
\end{equation}

\subsection{$SO(3)$ synchronization}
\label{sec:so3_synch_methods}
$SO(3)$ is the group of 3-D rotations. 
Each element of $SO(3)$ can be represented by a $3\times3$ matrix $R_i$ that satisfies $\det(R_i) =1$, and $R_iR_i^T=R_i^TR_i=I$, where $I$ is the identity matrix. 
The $SO(3)$  synchronization  problem is to estimate the block matrix
\begin{equation} \label{eq:R}
R=[R_1^T,\ldots,R_N^T]^T \in \mathbb{R}^{3N\times 3}, 
\end{equation}
given the noisy pairwise ratios:
\begin{equation}
\label{so3_data}
H = \frac{\lambda}{N}RR^T + \frac{1}{\sqrt{3N}}W,
\end{equation}
where \(W\) is a symmetric matrix whose entries are distributed independently (up to symmetry) as \(\mathcal{N}(0,1)\), and \(\lambda>0\) denotes the SNR parameter. The problem can be associated with the maximum likelihood estimation problem~\cite{singer2011three}:
\begin{equation}
\label{cost_synchronization_SO3}
\max_{R}{R^THR},
\end{equation}
where \(R\in\mathbb{R}^{3N\times3}\) is of the form~\eqref{eq:R}, and each $3\times 3$ block \(R_i\) is in $SO(3)$.

\subsubsection{Spectral method}
\label{so_3_spectral_method}
Similarly to synchronization over $\mathbb{Z}/2$ and $U(1)$, we begin by 
computing the three leading eigenvectors of $H\in\mathbb{R}^{3N\times3N}$, which we denote by $\hat{R}_1,\hat{R}_2,\hat{R}_3.$ 
This method is typically called the  spectral method~\cite{singer2011three}, and we omit the details of the power iterations for simplicity.  
Then, we form a matrix $\hat{R}=[\hat{R}_1, \hat{R}_2, \hat{R}_3]\in\mathbb{R}^{3N\times3}$,  and finally each $3\times 3$ block of $\hat{R}$ is projected onto the nearest orthogonal matrix.
This projection, denoted by $\text{project}_{SO(3)}$, takes a $3\times 3$ matrix $M$, computes its SVD factorization $M = U\Sigma V^T$ and replaces the diagonal matrix $\Sigma$ by an identity matrix so that $\text{project}_{SO(3)}(M)=\pm UV^T$. The sign is chosen so that the determinant is one.

\subsubsection{Projected power method (PPM)}
The $(t+1)$-th iteration of the PPM reads:
\begin{equation}
\label{so_3_iteration}
    R^{(t+1)} = \text{project}_{SO(3)}(H R^{(t)}).
\end{equation}
To initialize the algorithm, we draw $N$, $3\times 3$ matrices whose entries are drawn i.i.d.\ from $\mathcal{N}(0,1)$, and then project each matrix  to the nearest orthogonal matrix as described above.

We did not implement the AMP algorithm for $SO(3)$ synchronization.

\subsection{Multi-reference alignment (MRA)}
We consider two MRA setups. In both cases, assuming the SNR is not too low, 
we first estimate the pairwise ratios between the group elements from the observations. Then, we estimate the group elements using a synchronization algorithm,  align the noisy observations, and average out the noise. 

\subsubsection{MRA over $\mathbb{Z}/2$}
\label{mra_z_2_existing_methods}
We assume to acquire $N$ measurements of the form
\begin{equation}
\label{antipodal_signals_model}
y_i = s_i x + \frac{1}{\lambda}\varepsilon_i, \quad  i=1,\ldots,N,
\end{equation}
where \(x,\varepsilon_i\in\mathbb{R}^L\), \(\varepsilon_i\sim \mathcal{N}(0,I)\) and \(s_i\in\{-1,1\}\). 
Our goal is to estimate  \(x\), up to a sign, from $y_1,\ldots,y_N,$ when $s_1,\ldots,s_N,$ are unknown. 

To estimate $x$, we first build   the pairwise ratio matrix by 
\begin{equation}
	\label{estimated_relative_measurements_z_2}
	H_{ij}=\frac{\lambda}{N}y_i^T y_j\approx \lambda s_is_j,
\end{equation} 
and then estimate the  group elements \(\{s_i\}_{i=1}^N\) 
 using  one of the existing methods for \(\mathbb{Z}/2\) synchronization described in Section~\ref{z_2_existing_methods}.
 Let $\hat{s}_1,\ldots,\hat{s}_N,$ be the estimated group elements. Then, the signal can  be  reconstructed by averaging 
\begin{equation}
\label{z_2_alignment}
    \hat{x} = \frac{1}{N} \sum_{i=1}^N \hat{s}_i y_i.
\end{equation}

We emphasize that, in contrast to the synchronization problem,  the error in~\eqref{estimated_relative_measurements_z_2} is not Gaussian anymore; in fact, the error is correlated:
\begin{equation}
\begin{aligned}
H_{ij}=&\frac{\lambda}{N}y_i^T y_j=
\frac{\lambda}{N}s_is_j\|x\|_2^2+w_{i,j},
\end{aligned}
\end{equation}
where $w_{i,j}=\frac{1}{N}(x^T(s_j\varepsilon_i + s_i\varepsilon_j)+\frac{1}{\lambda}\varepsilon_i^T\varepsilon_j)$. Note that
$\mathbb{E}[w_{i,j}w_{i,k}]=
\frac{s_j s_k}{N^2}
\mathbb{E}[
(x^T\varepsilon_i)^2
]\neq 0 $, where the expectation is taken with respect to the noise terms. 

\subsubsection{MRA over the group $\mathbb{Z}/L$ of  circular shifts} 
\label{mra_z_l_existing_methods}
Now, we consider a set of measurements  of the form
\begin{equation}
\label{1d_mra_model}
y_i = R_{s_i}x + \frac{1}{\lambda}\varepsilon_i \quad  i=1,\ldots,N,
\end{equation}
where \(x\in\mathbb{R}^L\) is sought signal, \(R_s\) is a circular shift operator, that is, $R_s(x)[i]=x[(i-s )\bmod L]$,  \(s\sim U[0,L-1]\), and \(\varepsilon_i\sim\mathcal{N}(0,I)\). 
We wish to estimate \(x\), up to a circular shift, from $y_1,\ldots,y_N$, when   $s_1,\ldots,s_N$ are unknown. 

To estimate the signal, we first estimate the pairwise ratio between the group elements (namely, the relative circular shift) by 
taking the maximum of the cross correlation between pairs of observations. 
This can be computed efficiently using the FFT algorithm by the relation 
\begin{equation}
\label{estimated_ratios_mra}
s_{ij}=\arg\max\mathcal{F}^{-1}(\mathcal{F}(y_i) \circ \mathcal{F}^*(y_j)),
\end{equation}
where $\mathcal{F}$ stands for the Fourier transform and $\circ$ is an element-wise multiplication.
Then, we construct the pairwise matrix:
\begin{equation}
\label{ratios_matrix_mra}
H_{ij} = \frac{\lambda}{N}e^{\iota 2 \pi \frac{s_{ij}}{L}},
\end{equation}
and  estimate the group elements using one of the existing methods for \(U(1)\) synchronization described in \ref{u_1_existing_methods}. 
We keep the normalization to be consistent with the scaling of the synchronization model when the pairwise ratios are given.
Let $\hat{s}_1,\ldots,\hat{s}_N,$ be the estimates of the group elements. 
The signal can then be estimated by alignment and averaging. 
\begin{equation}
    \hat{x} = \frac{1}{N} \sum_{i=1}^N R_{-\hat{s}_i} y_i
\end{equation}

Throughout this work, we assume that the SNR is high enough so that the group elements can be estimated to a reasonable accuracy. 
We mention that when the SNR is very low, the group elements cannot be estimated reliably, and thus the strategy described above will fail. 
Several methods were developed to estimate the signal in such low SNR environments without estimating the group elements,  see, for instance,~\cite{BispectrumInversion,abbe2018multireference,perry2019sample}.

\section{Unrolled algorithms for group synchronization}
\label{sec:unrolling}
\begin{figure}[h]
	\includegraphics[width=\linewidth]{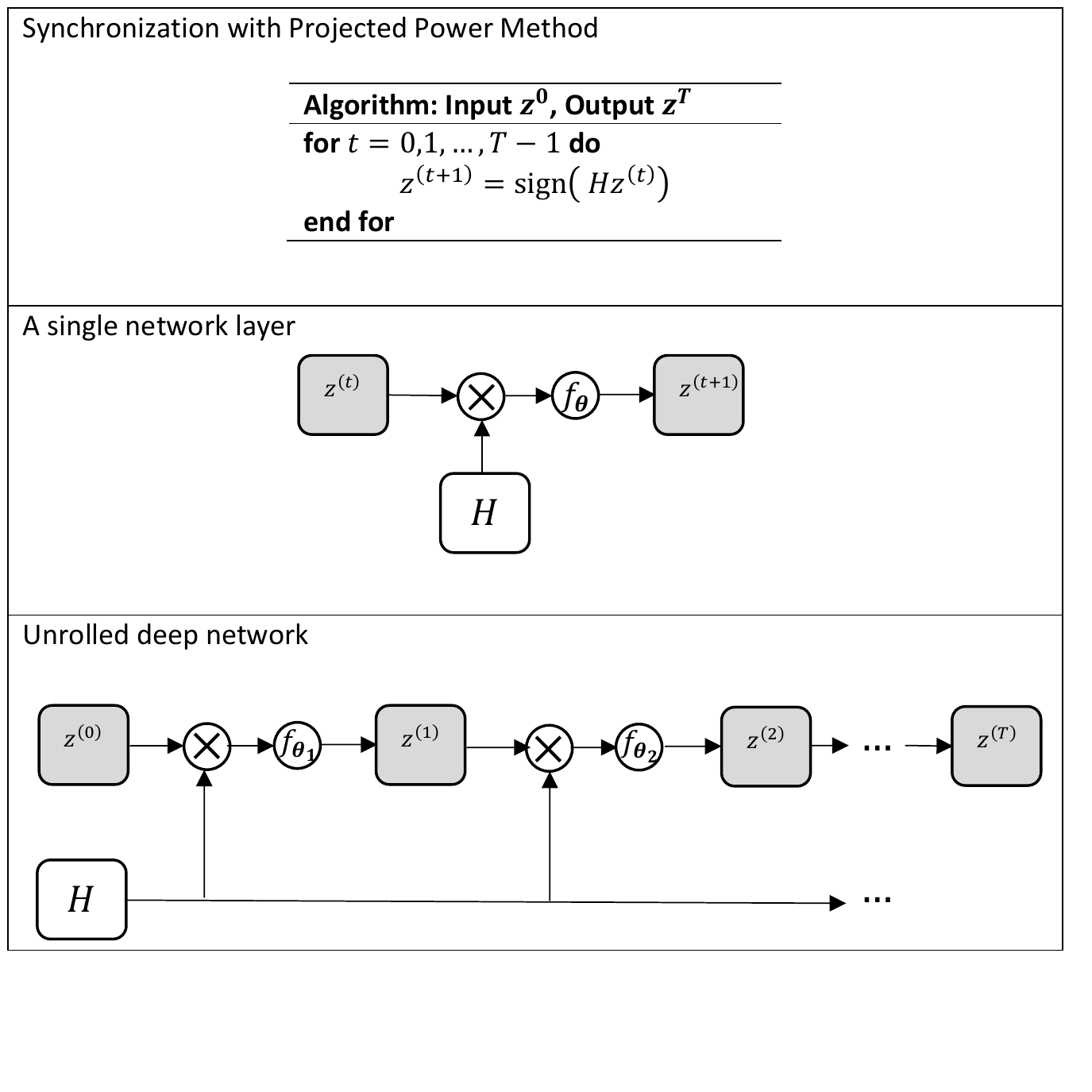}
	\caption{
	The concept of an unrolled algorithm  for \(\mathbb{Z}/2\) synchronization based on the projected power method. 
	The upper panel shows the standard projected power method for \(\mathbb{Z}/2\) synchronization with $T$ iterations; see Section~\ref{sec:sync_z2_ppm}. The middle panel illustrates a single iteration of the algorithm in the form of a single-layer network, where the sign operator is replaced by a general, learned non-linear function $f_\theta$. 
	The bottom panel shows the concatenation of $T$ layers into an unrolled deep network. Each layer may have a separate set of parameters.	
		The Onsager correction term that uses \(z^{(t-1)}\) is omitted for simplicity.  
		}
	\label{z_2_unrolling_concept}
\end{figure}

Based on the structural similarity between the group synchronization algorithms described in Section~\ref{sec:existing_methods} and deep neural networks, we adopt the concept of algorithm unrolling:  mapping each iteration of an iterative algorithm into a learned network layer, and stacking the layers together to form a deep neural network.
Each layer consists of multiplying  the current estimate of  group elements with the measurement matrix, \(Hz^{(t)}\), as in the iteration formula, but replaces the explicit non-linear function by a learned non-linear function. Each layer has the flexibility to incorporate information from the $(t-1)$-th layer.  
 Specifically, the $(t+1)$-th layer receives as an input the measurement matrix \(H\) and the previous estimates \(z^{(t)}\) and \(z^{(t-1)}\), and is parameterized by a set of weights \({\theta}^{(t)}\):
 \begin{equation}
 	z^{(t+1)}=\ell_{\theta^{(t)}}(z^{(t)},z^{(t-1)},H),
 \end{equation}
where $\ell$ denotes the architecture of the layer.
The layers can either share weights or have different weights per layer.
 Figure~\ref{z_2_unrolling_concept} illustrates the concept of an unrolled algorithm for $\mathbb{Z}/2$ synchronization. 
 
 In order to train the network, we generate data according to the data generative model, including the relative measurement matrix and the ground truth group elements. The network is trained using stochastic gradient descent to minimize a loss function that measures an error metric (up to a group symmetry) over a  batch of samples. 
 Thus, given an initial estimate \(z^{(0)}\), we get an estimator for the group elements of the form:
 \begin{equation}
 	\hat{z}=F_{\Theta}(z^{(0)},H),
 \end{equation}
 where \(\Theta\) is the entire set of weights: \(\Theta = [{\theta^{(0)}},...,{\theta^{(T-1)}}]\),  and \(F\) is the deep neural network function.

\revv{
	While we cannot provide theoretical guarantees, we  conjecture that the unrolling algorithm outperforms existing algorithms for the following reasons:
    \begin{enumerate}

        \item Existing solutions are not necessarily optimal and the error guarantees are for asymptotic settings, whereas we examine the finite-dimensional setting.

        \item The analysis of previous algorithms assumes that the errors of the relative group ratios are independent.
        However, usually the relative group ratios are estimated from the data (e.g., in cryo-EM), and thus this error model does not hold.  
        
        \item The starting point of this work was the resemblance of existing iterative synchronization algorithms to the blueprint of neural networks. We chose to use algorithm unrolling and not a generic neural network architecture to benefit from the advantages of algorithm unrolling: interpretable structure, which contains domain knowledge and require  less training data.
        
    \end{enumerate}
}
In the following subsections, we elaborate on  specific network architectures, including the loss functions, for  the models introduced in Section~\ref{sec:existing_methods}.

\subsection{Architecture and loss function for $\mathbb{Z}/2$ synchronization} \label{sec:Architecture_z2}
Following the AMP iterations in \eqref{z_2_amp_1} and \eqref{z_2_amp_2}, the $(t+1)$-th layer receives as input the measurement matrix \(H\in\mathbb{R}^{N\times N}\), and the previous layers' estimates \(z^{(t)},z^{(t-1)}\in\mathbb{R}^{N}\). The  output \(z^{(t+1)}\in\mathbb{R}^{N}\) can be described using the following equations: 
\begin{equation}
c = \theta_0\lambda Hz^{(t)} -\lambda^2(1-\langle (\phi_{\boldsymbol{\theta}_2}(z^{(t)}))^2\rangle)z^{(t-1)},
\end{equation}
and 
\begin{equation}
z^{(t+1)} = f_{\boldsymbol{\theta}_1}\left(c\right),
\end{equation}
where \(f\) and \(\phi\) are learned functions parameterized by a set of weights \(\boldsymbol{\theta}_1\) and \(\boldsymbol{\theta}_2\). 
 
We denote by Dense(N)  a linear layer with $N$ neurons, whose input is the previous layer's output, BatchNorm() denotes a batch normalization layer, ReLU() is a relu layer, and tanh() is a hyperbolic tangent  layer.
The learned function has the following structure: Dense(32) \(\to\) BatchNorm() \(\to\) ReLU() \(\to\) Dense(1) \(\to\) BatchNorm() \(\to\) tanh(), such that its outputs are in the range \([-1,1]\).

Given a batch of size \(M\), with ground truth and predicted group elements \(\{z_m\}_{m=1}^{M}\) and \(\{\hat{z}_m\}_{m=1}^{M}\), respectively, we use the following loss function to optimize  the weights:
\begin{equation} \label{eq:loss_z2}
\mathcal{L}(\Theta) = 
1 - \frac{1}{NM}\sum_{m=1}^{M}{
	|z_m^T \hat{z}_m|
},
\end{equation}
where $\Theta = [{\boldsymbol{\theta}^{(0)}},...,{\boldsymbol{\theta}^{(T-1)}}]$ is the set of parameters of the network, and $\boldsymbol{\theta}^{(t)}=[\theta^{(t)}_0, \boldsymbol{\theta}^{(t)}_1,\boldsymbol{\theta}^{(t)}_2]$ is the set of parameters per layer. The loss function~\eqref{eq:loss_z2} measures the average alignment error, up to a sign, between the predicted and the ground truth group elements   over $M$ samples. The absolute value function is required due to the sign symmetry.

\subsection{Architecture and loss function for $U(1)$ synchronization} \label{sec:ArchitectureU1}
Based on the AMP iterations \eqref{u_1_amp_1} and \eqref{u_1_amp_2}, the $(t+1)$-th layer receives as input \(H_r,H_i\in\mathbb{R}^{N\times N}\), the real and imaginary parts of the measurement matrix,  respectively, and  \(z^{(t)}_r,z^{(t)}_i,z^{(t-1)}_r,z^{(t-1)}_i\in\mathbb{R}^{N}\):  the real and imaginary parts of the estimates of the previous layers. The  output $z^{(t+1)}_r,z^{(t+1)}_i\in\mathbb{R}^{N}$ can be described using the following equations: 
\begin{align}
&c_r = \theta_0\lambda (H_r z^{(t)}_r - H_i z^{(t)}_i) -\lambda^2(1-\langle {z^{(t)}_r}^2+{z^{(t)}_i}^2\rangle)z^{(t-1)}_r, \nonumber \\
&c_i = \theta_0\lambda (H_r z^{(t)}_i + H_i z^{(t)}_r)-\lambda^2(1-\langle {z^{(t)}_r}^2+{z^{(t)}_i}^2\rangle)z^{(t-1)}_i	
\end{align}
and 
\begin{align}
&z^{(t+1)}_r[n] = \frac{c_r[j]}{\max\left(|c[j]|,\varepsilon\right)}f_{\boldsymbol{\theta}_1}\left(|c[j]|\right), 
\nonumber \\
&z^{(t+1)}_i[j] = \frac{c_i[j]}{\max\left(|c[j]|,\varepsilon\right)}f_{\boldsymbol{\theta}_1}\left(|c[j]|\right),
\end{align}
where $ |c[j]| = \sqrt{{c_r[j]}^2+{c_i[j]}^2}$, and 
 \(\varepsilon=10^{-12}\) is a small constant that is introduced for numerical stability. The non-linear function \(f\) is a learned function parameterized by a set of weights \(\boldsymbol{\theta}_1\) with the following structure: Dense(256) \(\to\) ReLU() \(\to\) Dense(1) \(\to\) tanh(), such that its outputs are within \([-1,1]\).

Let \(\{z_{r_m}\}_{m=1}^{M}, \{z_{i_m}\}_{m=1}^{M}\) and \(\{\hat{z}_{r_m}\}_{m=1}^{M}, \{\hat{z}_{i_m}\}_{m=1}^{M}\) be the real and imaginary parts of the ground truth and the predicted group elements, respectively, 
of a batch of size \(M\). We use the following loss function to optimize the weights:

\begin{equation}
\label{alignment_loss_u_1}
\begin{split}
\mathcal{L}(\Theta) = 
1 - \frac{1}{NM}\sum_{m=1}^{M}
	& 
	\left[
	(z_{r_m}^T \hat{z}_{r_m}+z_{i_m}^T \hat{z}_{i_m})^2\right.
	\\
	&
	\left.+(z_{r_m}^T \hat{z}_{i_m}-z_{i_m}^T \hat{z}_{r_m})^2\right]^{1/2},
\end{split}
\end{equation}
where $\Theta = [{\boldsymbol{\theta}^{(0)}},...,{\boldsymbol{\theta}^{(T-1)}}]$ is the set of network's parameters, and $\boldsymbol{\theta}^{(t)}=[\theta^{(t)}_0, \boldsymbol{\theta}^{(t)}_1]$. This loss function measures the alignment between the ground truth and predicted group elements, and it is invariant to  a global phase shift (the  symmetry of the problem).

\subsection{Architecture and loss function for $SO(3)$ synchronization}

The projection operation in equation~\eqref{so_3_iteration}, which consists of SVD factorization, is non-differentiable and thus gradients cannot be back-propagated through it during the learning process. 
Therefore, in order to unroll the projected power method into a differentiable neural network, this projection operation should be replaced. 
To derive a differentiable projection operation, we start with an alternative method that
expresses the nearest orthogonal matrix of a matrix \(A\), denoted by $Q$,  explicitly using the matrix square root: $Q=A(A^TA)^{-\frac{1}{2}}$.
This method can be combined with the Babylonian method, and a first order approximation suggests the following iterations after setting $Q_0=A/\|A\|_{\text{F}}$~\cite{orthogonalmatrix}: 
\begin{equation} \label{eq:babylonian}
\begin{aligned} 
    & N_i = Q_i^T Q_i \\
    & P_i = \frac{1}{2}Q_i N_i \\
    & Q_{i+1} = 2 Q_i + P_i N_i - 3 P_i. 
\end{aligned}
\end{equation}
Numerical experiments suggest that this recursion typically converges after 4 iterations. 
We thus use \(Q_4\) as an estimation for the nearest orthogonal matrix of $A$,  through which gradients can be backpropagated.

The unrolled synchronization algorithm for $SO(3)$ is composed of a stacked learned synchronization blocks, followed by a projection block as the last layer. 
Each learned synchronization block takes on the form:
\begin{equation}
    R^{(t+1)} = f_{\boldsymbol{\theta}_1}(HR^{(t)}) + \phi_{\boldsymbol{\theta}_2}(R^{(t-1)}).
\end{equation}
The function implementation consists of the following layers:
Reshape input $(M,3N,3)$ to $(M,N,9)$ $\to$ Dense(hidden~neurons) $\to$ BatchNorm() $\to$ ReLU() $\to$ Dense(9) $\to$ BatchNorm() $\to$ tanh() $\to$ Reshape into $(M,3N,3)$,
where $M$ is the batch size.
The first layer reshapes the input such that each $3\times 3$ block is flattened into 9 elements, resulting in a shape of $(M,N,9)$. The following layers apply the same non-linear functions to each 9-element vector and reshape them back into the dimensions of the input.
The function $f$ uses 32 hidden neurons and $\phi$ uses 9 hidden neurons. 
The function $\phi$ acts as the Onsager correction term and slightly improves the results.

The implementation of the projection block is as follows:
\begin{enumerate}
    \item reshape input $(M,3N,3)$ to $(M,N,3,3)$;
    \item normalize each  $3\times 3$ matrix by its Frobenius norm and apply the four iterations of~\eqref{eq:babylonian};   
    \item reshape the output of the last stage to $(M,3N,3)$.
\end{enumerate}

Given a batch of samples of size \(M\), with ground truth and predicted group elements \(\{R_m\}_{m=1}^{M}\) and \(\{\hat{R}_m\}_{m=1}^{M}\), respectively, we use the following loss function  to optimize  the weights:
\begin{equation}
\label{so3_loss}
\mathcal{L}(\Theta) = 
1 - \frac{3}{NM}\sum_{m=1}^{M}{
	\|R_m^T\hat{R}_m\|_{\text{F}}^2
},
\end{equation}
where $\Theta = [{\boldsymbol{\theta}^{(0)}},...,{\boldsymbol{\theta}^{(T-1)}}]$ is the set of network's parameters, and $\boldsymbol{\theta}^{(t)}=[\boldsymbol{\theta}^{(t)}_1,\boldsymbol{\theta}^{(t)}_2]$. The suggested loss measures the alignment between the ground truth and the predicted group element matrices,  and is invariant under a global rotation.

\subsection{Multi-reference alignment (MRA)}
MRA models differ from group synchronization in two important aspects. First, the goal of the MRA problem is not to estimate the group elements, but the signal itself.
Second, the pairwise ratios are not directly available, and are estimated from the observations. 
Therefore, the learning phase of MRA models is slightly different from group synchronization, as described below.
We draw  $M$ signals from some distribution. Then, for each signal, we  generate $N$ noisy measurements according to the MRA statistical model, and estimate the pairwise ratio between the corresponding group elements. 
Given the pairwise ratio matrix, we solve a group synchronization problem and aim to estimate the signal itself, up to a group action. 
As we will see below, this process suggests different loss functions than the ones used for group synchronization. 

\subsubsection{MRA over $\mathbb{Z}/2$}
\label{sec:MRA_z_2_architecture}
A direct application of the \(\mathbb{Z}/2\) architecture described in Section~\ref{sec:Architecture_z2}, when the pairwise ratios are estimated from the noisy measurements, only leads to a small improvement, as will be presented in Section~\ref{sec:results}. 
Therefore, we suggest  to incorporate the measurements themselves in the loss function of the neural network.

Let \(Y_m\in \mathbb{R}^{L\times N}\) be the measurement matrix of the \(m\)-th signal  \(x_m\in \mathbb{R}^{L}\), so that $Y_m[:,n]\in\mathbb{R}^{L}$ is the $n$-th observation of the $m$-th signal.
We suggest  the following reconstruction loss:
\begin{equation}
\label{reconstruction_loss}
\mathcal{L}_R(\Theta)=
\frac{1}{LM}\sum_{m=1}^{M}{
	\min_{s\in\{-1,1\}}\left\|x_m - \frac{s}{N}\sum_{n=1}^N{Y_m[:,n]\hat{z}_m[n]}\right\|^2
},
\end{equation}
where 
 \(\hat{z}_m\in \mathbb{R}^{N}\) is the predicted group elements output of the network  described in Section~\ref{sec:Architecture_z2}. 
 The loss function depends on the parameters $\Theta$ through the group elements $\{\hat{z}_m[n]\}_{n,m=1}^{N,M}$.
 Note that the reconstruction loss is invariant to the inherent sign symmetry.

\subsubsection{MRA over the group $\mathbb{Z}/L$ of circular shifts}
\label{sec:MRA_z_l_architecture}
Similarly to the  MRA model described above, 
when the relative shifts were estimated from the MRA  measurements, 
only a minor improvement in signal estimation was achieved using the architecture of 
$U(1)$ synchronization from Section~\ref{sec:ArchitectureU1}.
Thus, we aim to work with the measurements directly. 

It is more convenient to express the loss function in Fourier domain, where a circular shift is mapped to a complex exponential.
Let \(\mathcal{X}_m\in\mathbb{C}^{L}\) be the Fourier transform of the $m$-th signal, and let \(\mathcal{Y}_m\in\mathbb{C}^{L\times N}\)
be the corresponding measurement matrix, where $\mathcal{{Y}}_{m}[:,n]$ is the Fourier transform of the $n$-th measurement of the $m$-th signal. 
 Let $\mathcal{Y}_{r_m}$ and $\mathcal{Y}_{i_m}$ denote the real and imaginary parts of $\mathcal{Y}_{m}$, and let $\hat{z}_m\in U(1)$ be the estimated rotation
 using the synchronization algorithm described in Section~\ref{sec:ArchitectureU1}. Note that $\hat{z}_m$ lies on the unit circle, whereas the circular shifts are discrete. 
 The real and imaginary parts of the aligned data matrix of the  $m$-th sample can be written as: 
\begin{align}
\label{fourier_alignment}
\mathcal{\tilde{Y}}_{r_m}[k,n] &= 
\cos\left(k\angle{\hat{z}_m}\right) \mathcal{Y}_{r_m}[k,n]- \sin\left(k\angle{\hat{z}_m}\right) \mathcal{Y}_{i_m}[k,n] \nonumber \\ 
\mathcal{\tilde{Y}}_{i_m}[k,n] &= 
\cos\left(k\angle{\hat{z}_m}^T\right) \mathcal{Y}_{i_m}[k,n] + \sin\left(k\angle{\hat{z}_m}^T\right) \mathcal{Y}_{r_m}[k,n],
\end{align}
for \(k=0,...,L-1\). The signal is then estimated by averaging:
\begin{equation}
\label{fourier_average}
\hat{\mathcal{X}}_m = \frac{1}{N}\sum_{n=1}^{N}{\left(\mathcal{\tilde{Y}}_{r_m}[:,n]+\I\mathcal{\tilde{Y}}_{i_m}[:,n]\right)}.
\end{equation}
Therefore, we use the following  loss function:
\rev{
\begin{equation}
	\label{reconstruction_loss_1d_mra}
	\mathcal{L}_R(\Theta) = 
	c\sum_{m=1}^{M}{
		\min_{\phi\in\Phi_P}\sum_k\left(\mathcal{X}_m[k] - e^{jk\phi}\mathcal{\hat{X}}_m[k]\right)^2,
	}
\end{equation}}
\rev{where $c=\frac{1}{L^2M}$ and  $\Phi_P=\{\frac{2\pi}{LP},2\frac{2\pi}{LP},\ldots,2\pi\}$.} 
In the numerical experiments below, we set \(P=10\).

\section{Numerical experiments}
\label{sec:results}

The following experiments  examine the average  error of the unrolled algorithms 
and the iterative algorithms described in Section~\ref{sec:existing_methods}.
In all experiments, we set $N=20$, and the number of test samples is equal to the number of training samples.
The code to reproduce all experiments  is publicly available at \url{https://github.com/noamjanco/unrolling_synchronization}.

\subsection{$\mathbb{Z}/2$ synchronization}
For a vector of ground truth group elements \(z\in\{\pm1\}^N\) and a prediction \(\hat{z}\),  the alignment error is defined as:
\begin{equation}
\label{alignment_error_z_2}
\e(z,\hat{z}) = 1 - \frac{|z^T\hat{z}|}{N}.
\end{equation} 
We note that \(\e(z,\hat{z})=0\) for an ideal estimation, where $\hat{z}=\pm z$.
In addition, the error is invariant to a global sign, i.e., \(\e(z,-\hat{z})=\e(z,\hat{z})\).

Each observation of length $N=20$ was generated according to ~\eqref{z_2_gaussian_model}, where each entry was drawn i.i.d.\ from a uniform distribution over $\pm 1$. 
The network was trained using a dataset of size  \(M = 20000\), with \(300\) epochs, and a learning rate of $10^{-3}$, using the Adam optimizer with batch size of $128$.

\begin{figure*}
	\begin{subfigure}[ht]{.66\columnwidth}
		\centering
		\includegraphics[width=\columnwidth]{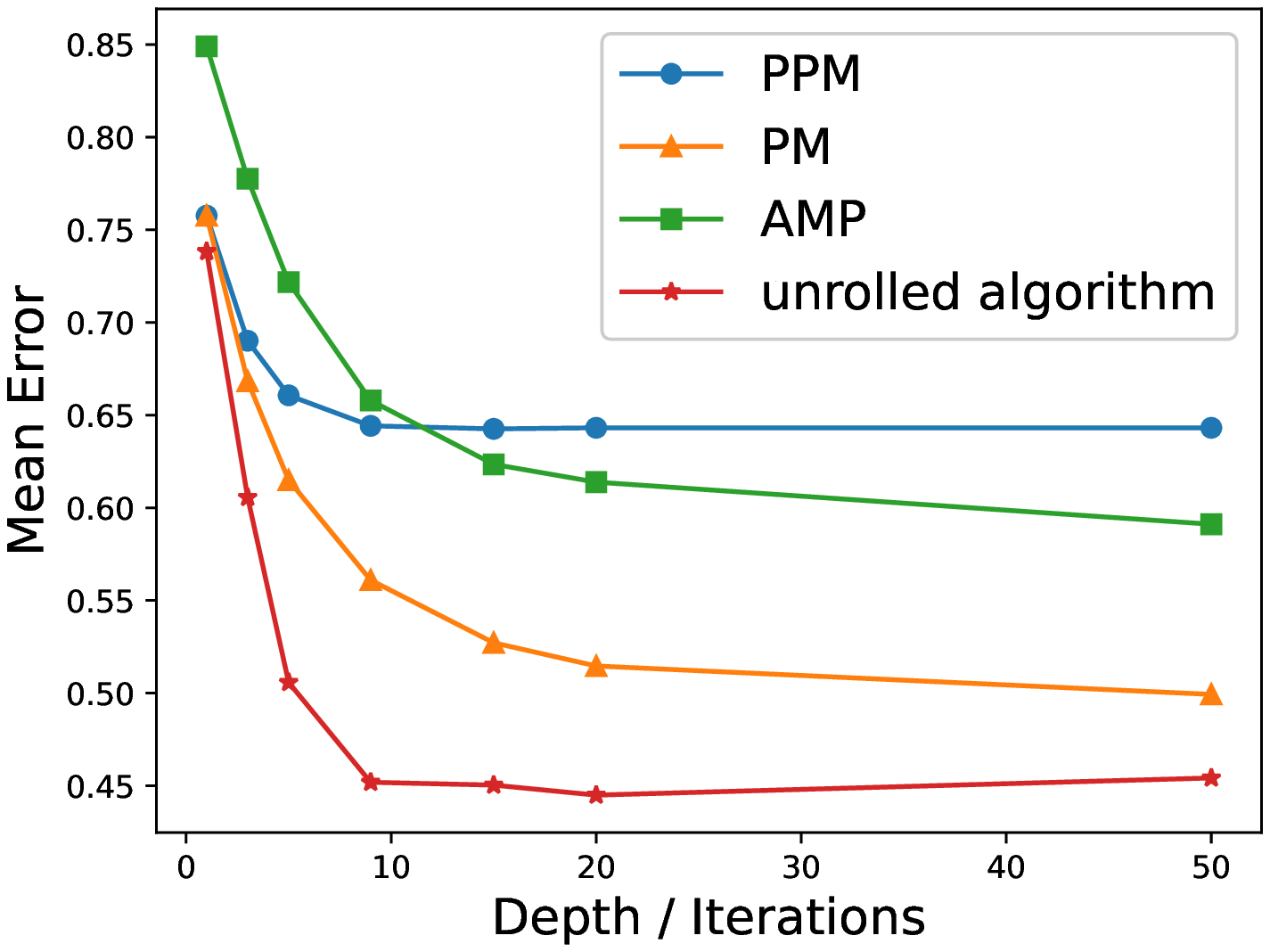}
		\caption{\(\lambda=1.2\)}
	\end{subfigure}
	\hfill
	\begin{subfigure}[ht]{.66\columnwidth}
		\centering
		\includegraphics[width=\columnwidth]{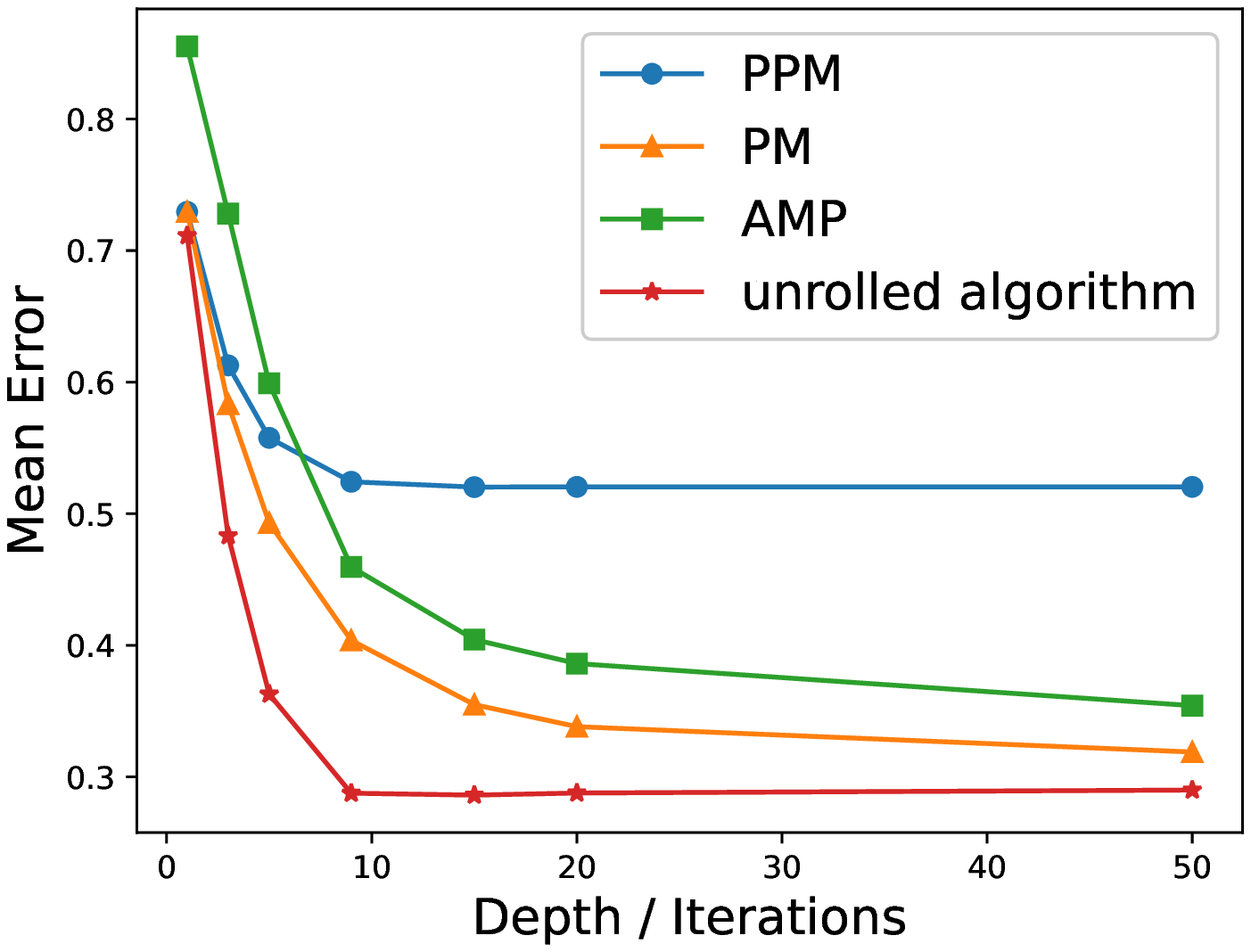}
		\caption{\(\lambda=1.5\)}
	\end{subfigure}
	\hfill
\begin{subfigure}[ht]{.66\columnwidth}
	\centering
	\includegraphics[width=\columnwidth]{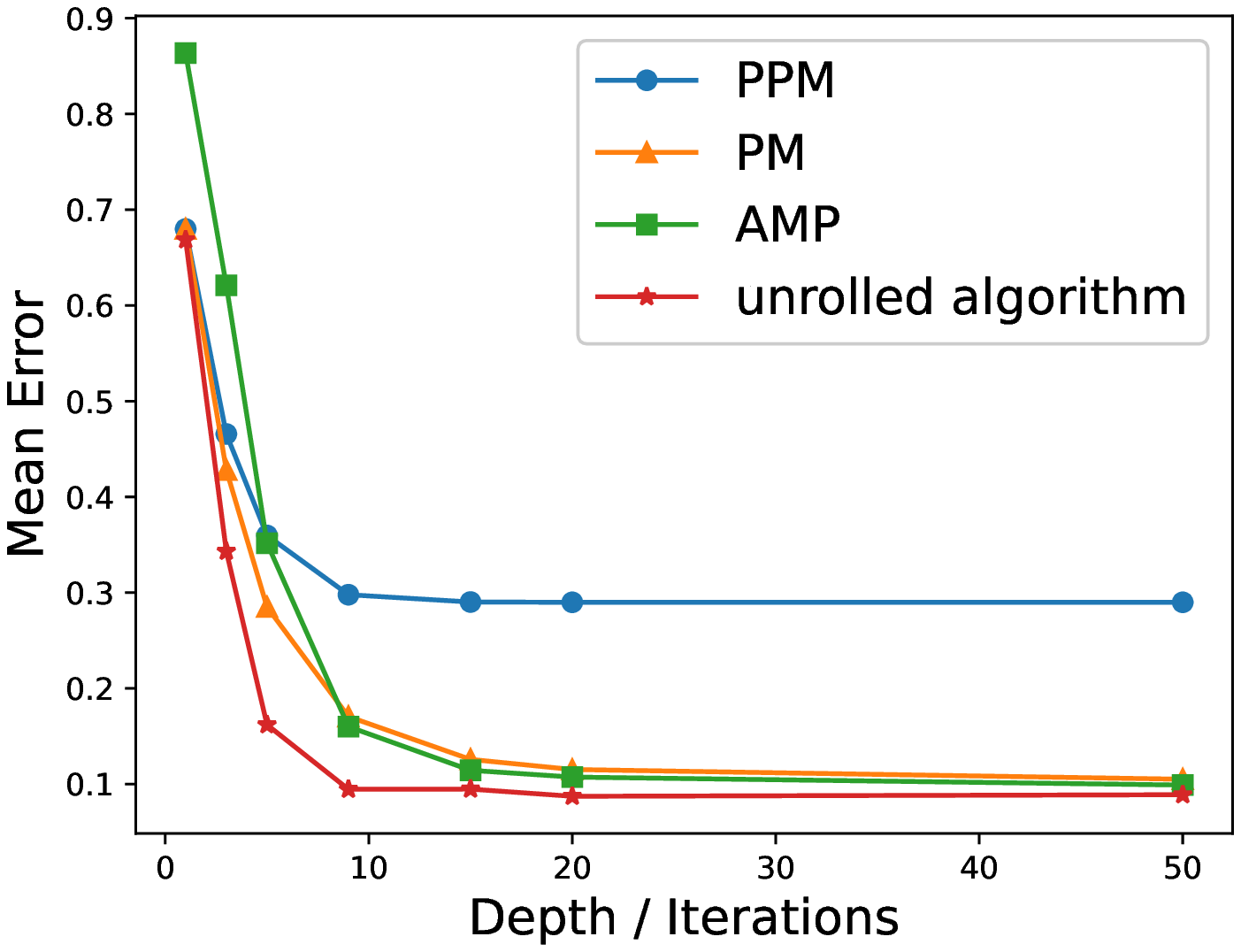}
		\caption{\(\lambda=2\)}
	\end{subfigure}
	\caption{\label{results_z_2} Alignment error \eqref{alignment_error_z_2} as a function of  depth for the $\mathbb{Z}/2$ synchronization problem with different \(\lambda\) values. The unrolled algorithm is compared against the power method (PM), projected power method (PPM), and the AMP algorithm described in Section~\ref{z_2_existing_methods}.
	The unrolled synchronization network outperforms the alternative algorithms, and the gap increases as the SNR decreases. 
		}
\end{figure*}

The alignment error as a function of depth is presented in Figure~\ref{results_z_2}, for SNR values of \(\lambda=1.2\), \(\lambda=1.5\) and \(\lambda=2\). 
We compared the performance of the unrolled algorithm against the alternative algorithms described in Section~\ref{z_2_existing_methods}, where the number of iterations is equal to the depth of the network. 
The results demonstrate that the unrolled synchronization network achieves better error performance, and the performance gap increases as the SNR decreases. 

\begin{figure}[h]
\includegraphics[width=\linewidth]{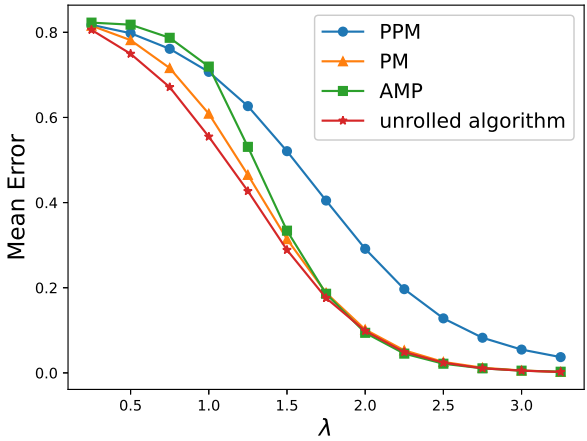}
	\caption{Alignment error \eqref{alignment_error_z_2} as a function of the SNR for $\mathbb{Z}/2$ synchronization. The depth of the unrolled algorithm is fixed to 9, while the alternative algorithms  ran for 100 iterations. Nevertheless,  the unrolled algorithm clearly  outperforms the iterative methods.
}
	\label{results_z_2_lambda_2_per_snr}
\end{figure}

Figure \ref{results_z_2_lambda_2_per_snr} shows the alignment error as a function of SNR, with a network of a fixed depth of 9, while the alternative algorithms used 100 iterations. We see that the neural network outperforms the alternative methods in terms of alignment error with much fewer iterations.

\subsection{$U(1)$ synchronization}
We define the error between the vector of the ground truth group elements \(z\in \mathbb{C}_1^N\) and a prediction \(\hat{z}\in \mathbb{C}_1^N\) by:
\begin{equation}
\label{alignment_error_u_1}
\e(z,\hat{z}) = 1 - \frac{|z^*\hat{z}|}{N}.
\end{equation} 
We note that \(\e(z,\hat{z})=0\) when $\hat{z} = ze^{\I\phi}$ for any $\phi\in[0,2\pi)$.
Generally,  the error is invariant to a global phase since \(\e(z,e^{\I\phi}\hat{z})=\e(z,\hat{z})\) for any $\phi\in [0,2\pi)$.

\begin{figure*}
	\begin{subfigure}[ht]{.66\columnwidth}
		\centering
		\includegraphics[width=\columnwidth]{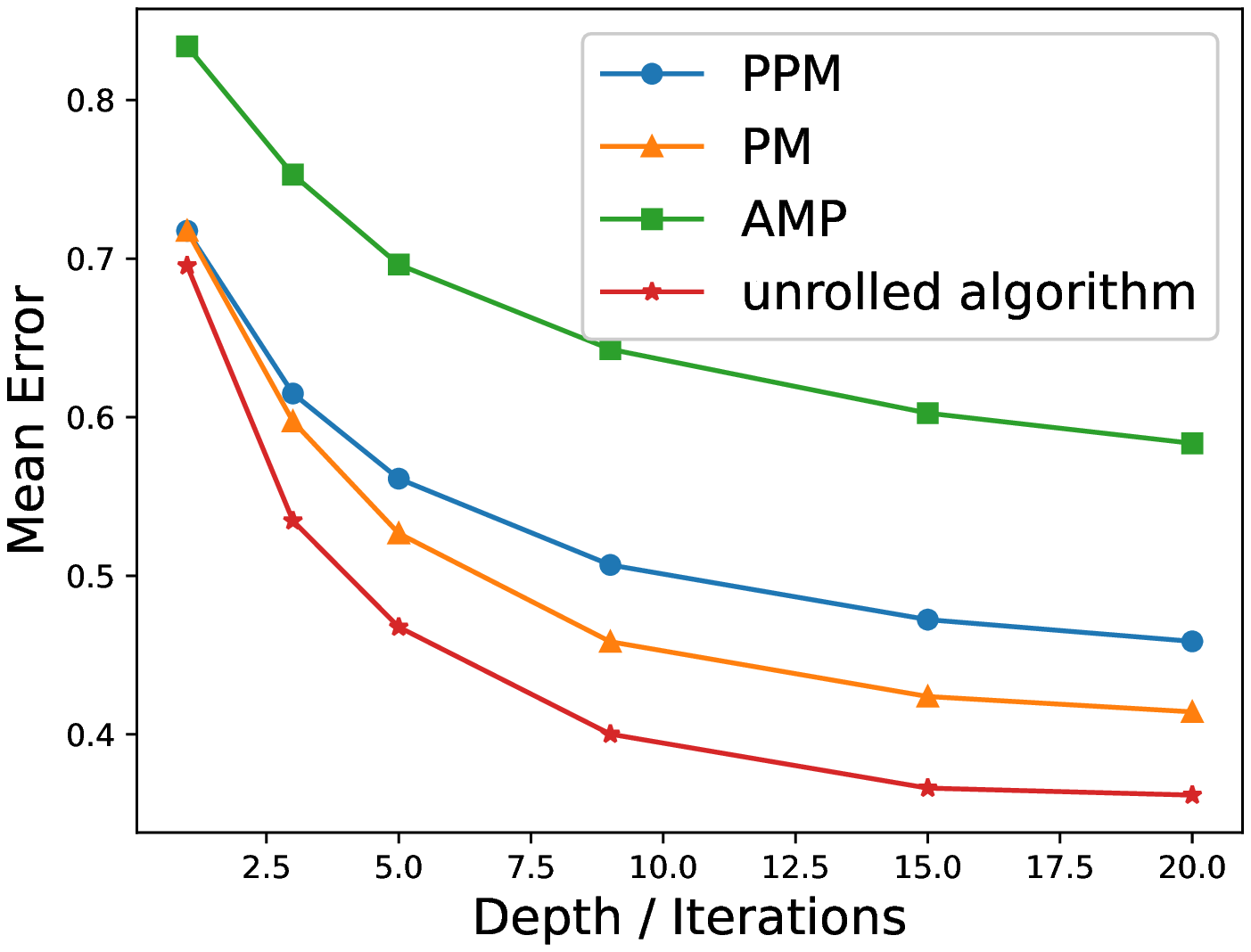}
		\caption{\(\lambda=1.2\)}
	\end{subfigure}
	\hfill
	\begin{subfigure}[ht]{.66\columnwidth}
		\centering
		\includegraphics[width=\columnwidth]{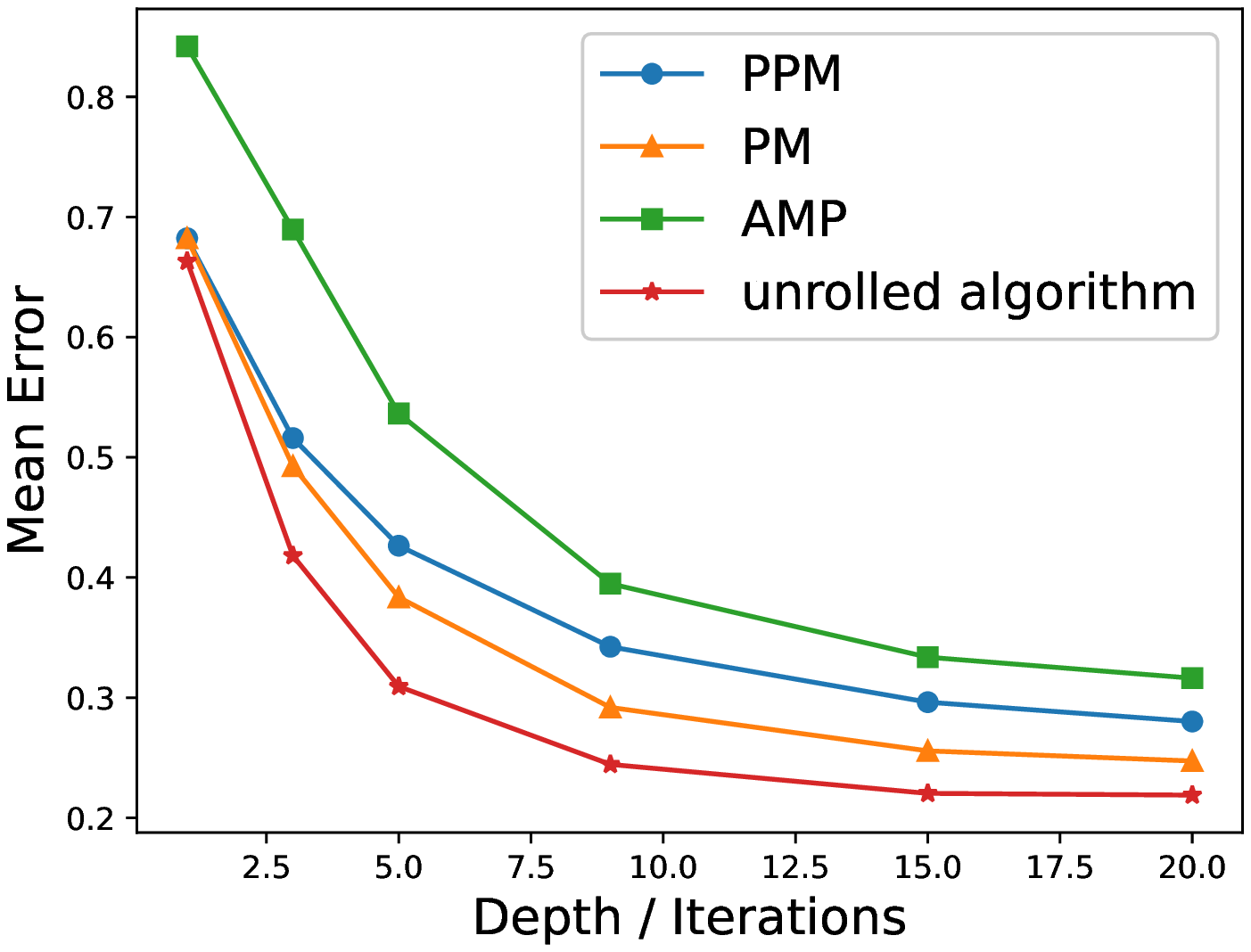}
		\caption{\(\lambda=1.5\)}
	\end{subfigure}
	\hfill
\begin{subfigure}[ht]{.66\columnwidth}
	\centering
	\includegraphics[width=\columnwidth]{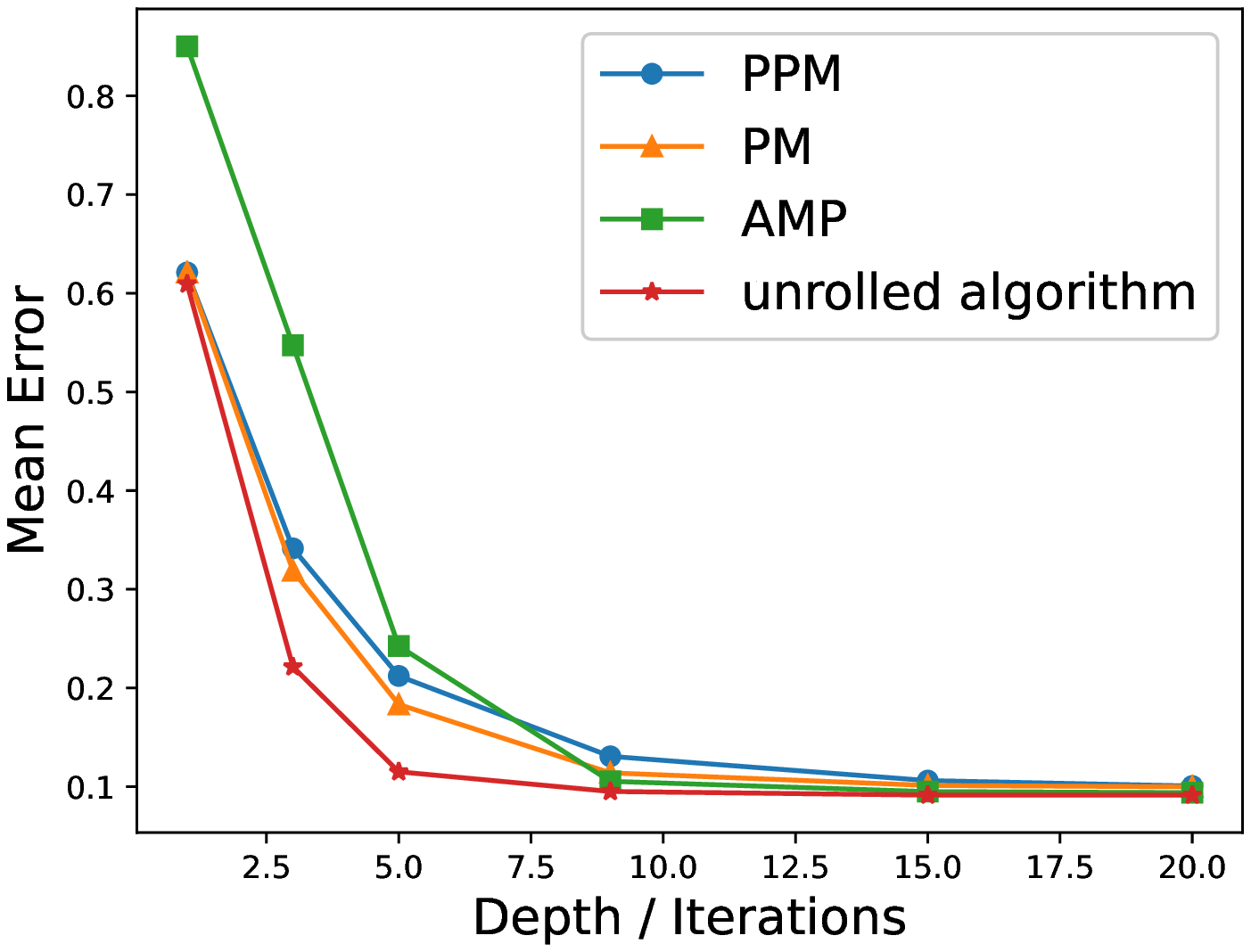}
		\caption{\(\lambda=2\)}
	\end{subfigure}
	\caption{\label{results_u_1} Alignment error \eqref{alignment_error_u_1} as a function of  depth for the $U(1)$ synchronization problem with different \(\lambda\) values. The unrolled algorithm is compared against the power method (PM), projected power method (PPM), and the AMP algorithm described in Section~\ref{u_1_existing_methods}.}
\end{figure*}

The network was trained using a data set of dimension $N=20$ and \(M=20000\) 
samples generated  according to the model in~\eqref{u_1_gaussian_model}. 
We used  the Adam optimizer with batch size of  $128$,  \(300\) epochs, and a learning rate of $10^{-4}$. The results are presented in Figure~\ref{results_u_1} for \(\lambda=1.2\), \(\lambda=1.5\) and \(\lambda=2\).
As in the $\mathbb{Z}/2$ synchronization,  the unrolled synchronization network outperforms the alternative algorithms, especially as the  SNR decreases.

\subsection{$SO(3)$ synchronization}
For a ground truth matrix \(R\in\mathbb{R}^{3N\times 3}\) (composed of $N, 3\times 3$ rotation matrices) and a prediction \(\hat{R}\in\mathbb{R}^{3N\times 3}\), the  error is defined as:
\begin{equation}
\label{alignment_error_so_3}
\e(R,\hat{R}) = 1 - \frac{3}{N}\left\|R^T\hat{R}\right\|_{\text{F}}^2.
\end{equation} 
This error metric is invariant to  a right multiplication by an orthogonal matrix,
and is 
equal to zero if $\hat{R}$ is equal to $R$  (up to a global rotation). 

\begin{figure*}
	\begin{subfigure}[ht]{.66\columnwidth}
		\centering
		\includegraphics[width=\columnwidth]{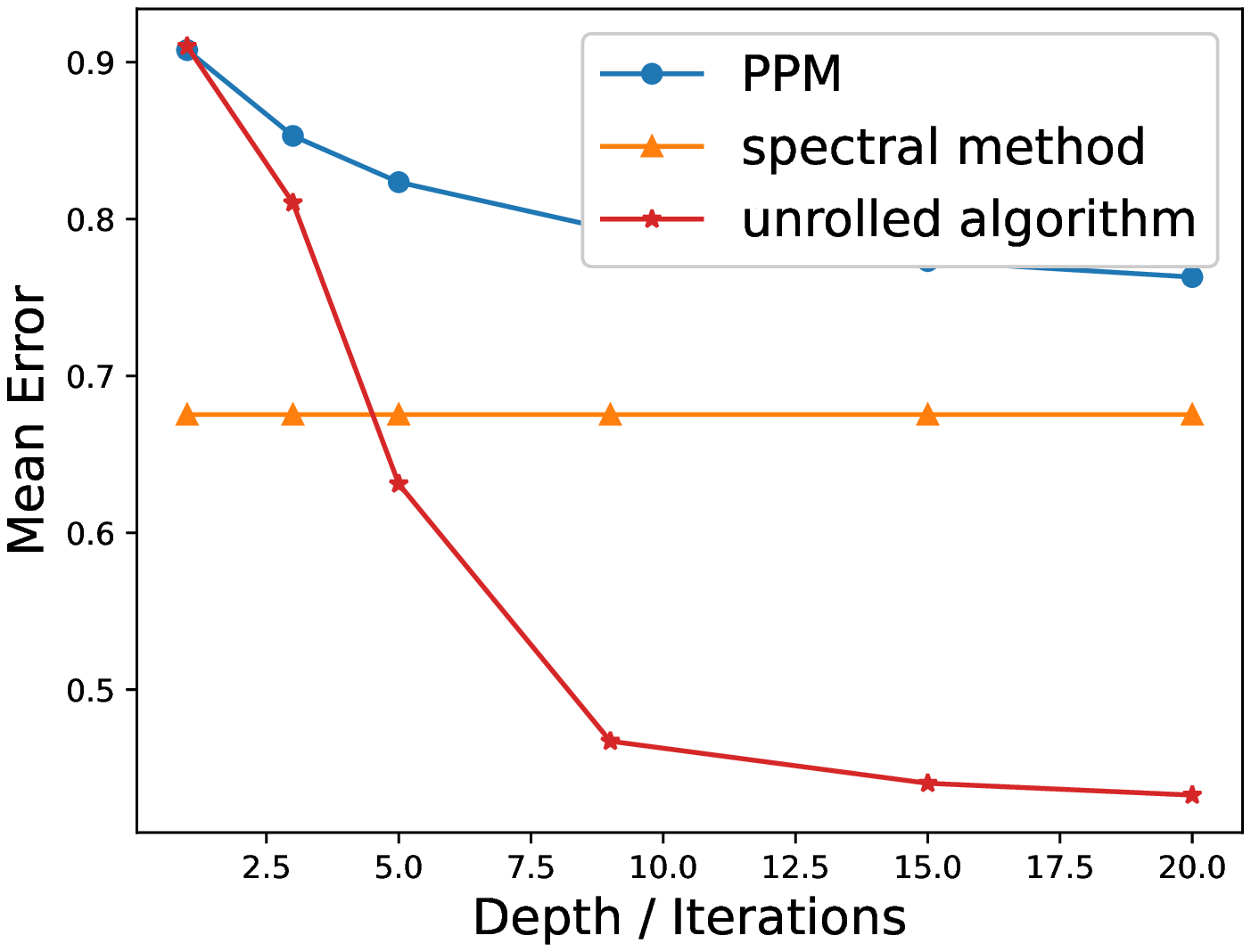}
		\caption{\(\lambda=1.2\)}
	\end{subfigure}
	\hfill
	\begin{subfigure}[ht]{.66\columnwidth}
		\centering
		\includegraphics[width=\columnwidth]{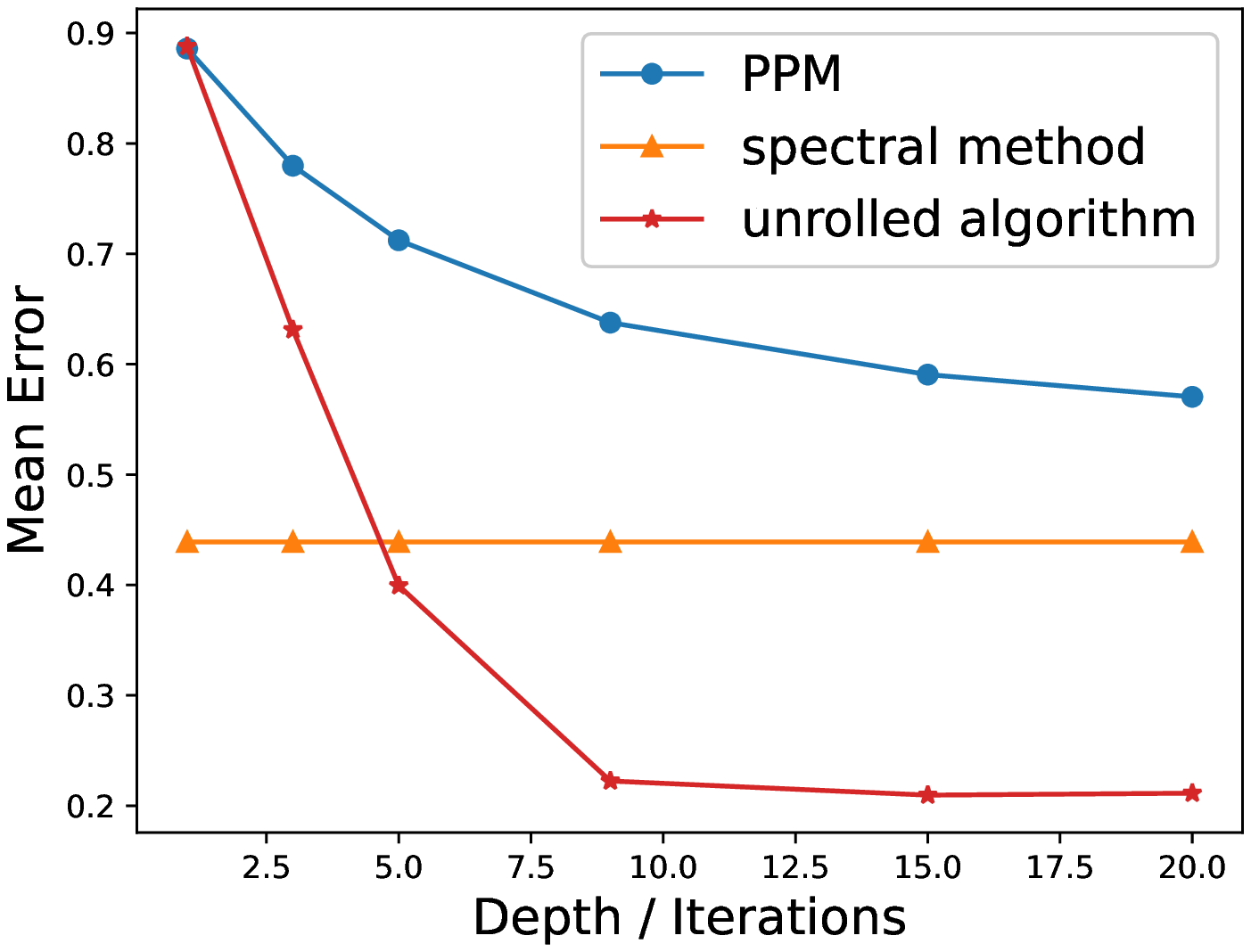}
		\caption{\(\lambda=1.5\)}
	\end{subfigure}
	\hfill
\begin{subfigure}[ht]{.66\columnwidth}
	\centering
	\includegraphics[width=\columnwidth]{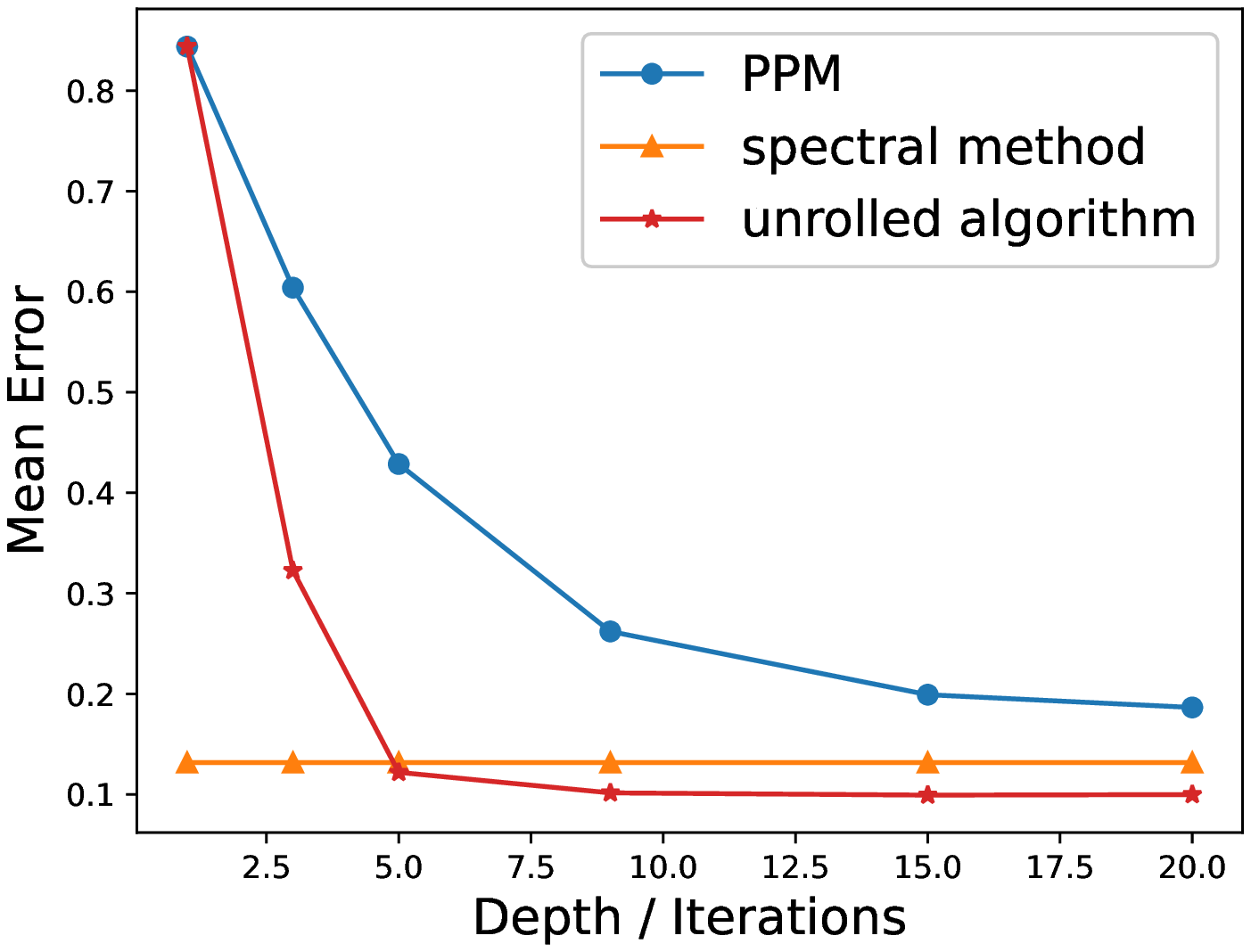}
		\caption{\(\lambda=2\)}
	\end{subfigure}
	\caption{\label{results_so_3} Alignment error \eqref{alignment_error_so_3} as a function of  depth for the $SO(3)$ synchronization problem with different \(\lambda\) values. The unrolled algorithm is compared against the spectral method and the projected power method (PPM) as described in Section~\ref{sec:so3_synch_methods}. 
We note that the spectral method computes the eigenvectors using SVD factorization, and thus the error is not a function of the number of iterations. 
}
\end{figure*}

The network was trained using a dataset  of dimension $N=20$ and \(M=10000\) samples generated according to the model in~\eqref{so3_data}. 
We used the Adam optimizer with batch size of $128$, \(300\) epochs, and a learning rate of $10^{-2}$. 
The spectral method computed the first three eigenvectors of the measurement matrix  using SVD factorization as described in \ref{so_3_spectral_method}. Therefore, its error is not a function of the number of iterations.  
The results are presented in Figure~\ref{results_so_3}, 
demonstrating a substantial gap between the unrolled algorithm and the competitors. 

\begin{figure}[h]
\includegraphics[width=\linewidth]{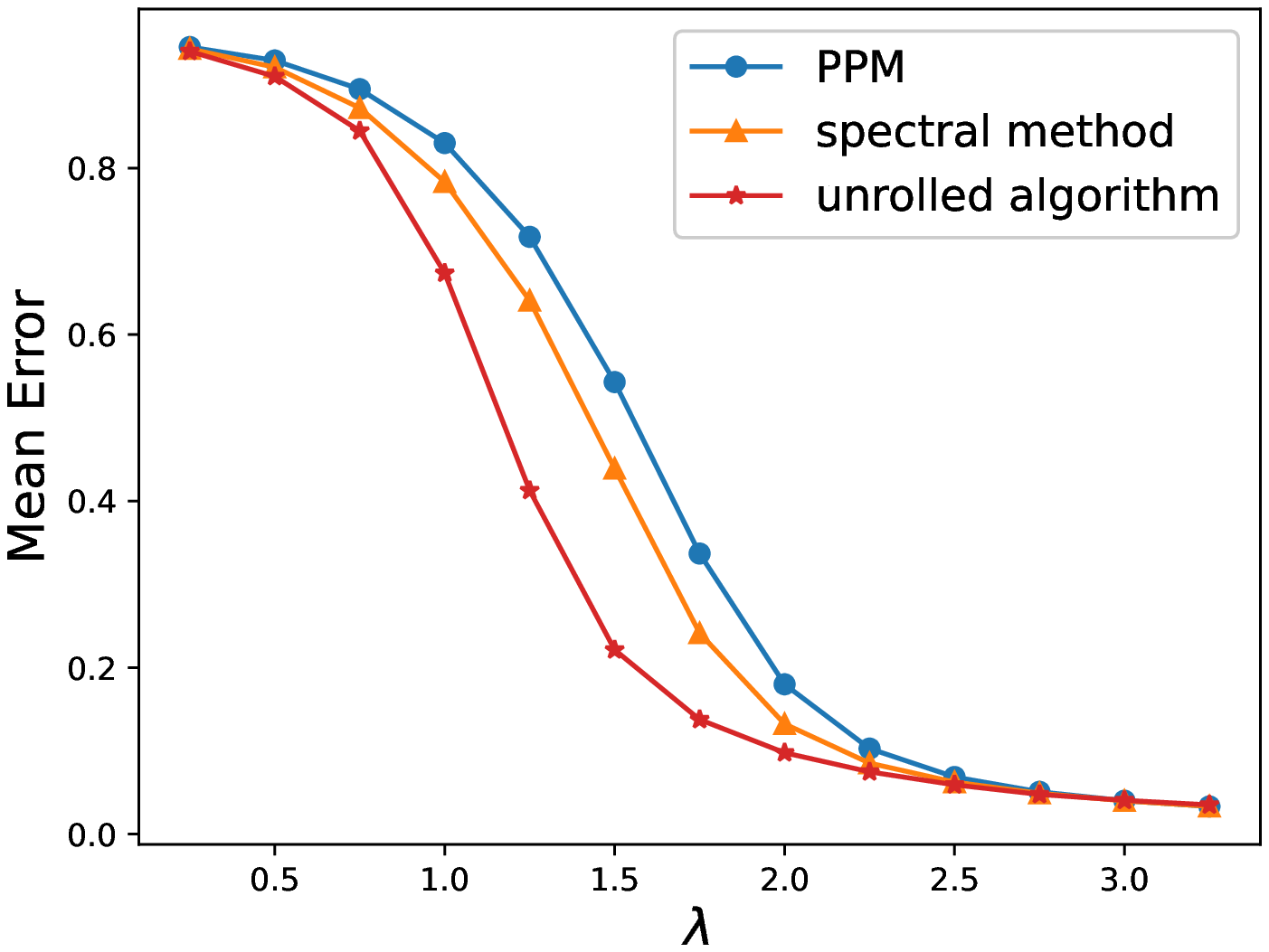}
	\caption{Alignment error \eqref{alignment_error_so_3} as a function of the SNR for $SO(3)$ synchronization. The depth of the unrolled algorithm is fixed to 9, while the alternative algorithms  ran for 100 iterations. Nevertheless,  the unrolled algorithm clearly  outperforms the iterative methods.}
	\label{results_so_3_per_snr}
\end{figure}

Figure~\ref{results_so_3_per_snr} shows the  error as a function of the SNR, 
when the depth of the network was fixed to 9, while the projected power method ran for 100 iterations. 
Nevertheless,  the unrolled algorithm clearly  outperforms the other methods.

\revv{
In addition, we measured the inference run-time of a batch of 10000 samples with  $\lambda=1.5$, $N=20$ and $L=9$ layers, and compared it against PPM with 100 iterations and the spectral method. The results are summarized in Table~\ref{tab:SO3_runtime}. The unrolled algorithm outperforms both methods in terms of alignment error and total run-time due to its low number of layers, with only a slight increase in run-time per layer.
    \begin{table}[h]
        \centering
        \resizebox{\columnwidth}{!}
        {
        \begin{tabular}{|c|c|c|c|}
        \hline
            Algorithm & Alignment Error & Total run-time [sec] & Single iteration run-time [sec]\\
        \hline
            Spectral method &  0.439003 & 18.96 & -\\
        \hline

            Projected power method & 0.637658 & 37.52 & \textbf{0.375}\\
        \hline

             Unrolled synchronization & \textbf{0.221980} & \textbf{3.53} & 0.392\\ 
             
        \hline
        \end{tabular}
        }
        \caption{Run-time for SO(3) synchronization with a batch of 10000 samples, $\lambda=1.5$, $N=20$ and $L=9$ layers, compared against PPM with 100 iterations and the spectral method.}
        \label{tab:SO3_runtime}
    \end{table}
}

\subsection{Multi-reference alignment over $\mathbb{Z}/2$}

We generated measurements according to \eqref{antipodal_signals_model} with a signal length of 21, where each entry was drawn  i.i.d.\ 
 from
$\mathcal{N}(0,1)$, and $N=20$. The relative ratios were estimated according to~\eqref{estimated_relative_measurements_z_2}. In the first part, we evaluate the alignment error \eqref{alignment_error_z_2} using the network described in \ref{sec:Architecture_z2}. In the next part, we evaluate the reconstruction error, defined as:
\begin{equation} \label{eq:rec_error_z2}
    \text{error}(x,\hat{x})=\min_{s\in\{-1,1\}}\left\|x - s\hat{x}\right\|^2,
\end{equation}
where $\hat{x}$ is the estimated signal, computed by aligning the measurements according to the estimated group elements and averaging, as described in \eqref{z_2_alignment}.
In this case, we used a modified loss function as described in~\ref{sec:MRA_z_2_architecture}.

\subsubsection{With alignment loss \eqref{eq:loss_z2}}
\begin{figure*}
	\begin{subfigure}[ht]{\columnwidth}
		\centering
		\includegraphics[width=\columnwidth]{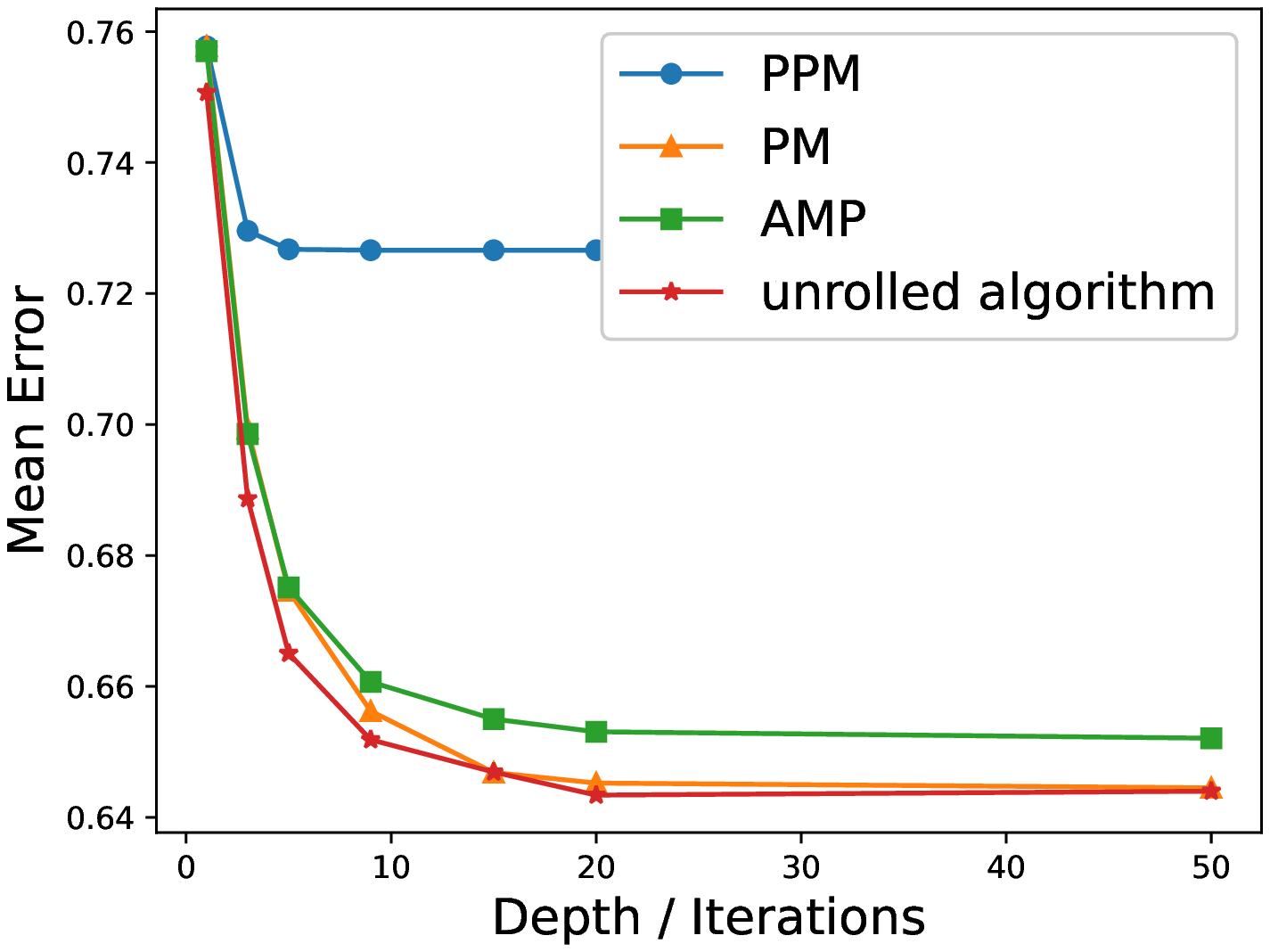}
		\caption{\(\lambda=0.2\)}
	\end{subfigure}
	\hfill
	\begin{subfigure}[ht]{\columnwidth}
		\centering
        \includegraphics[width=\columnwidth]{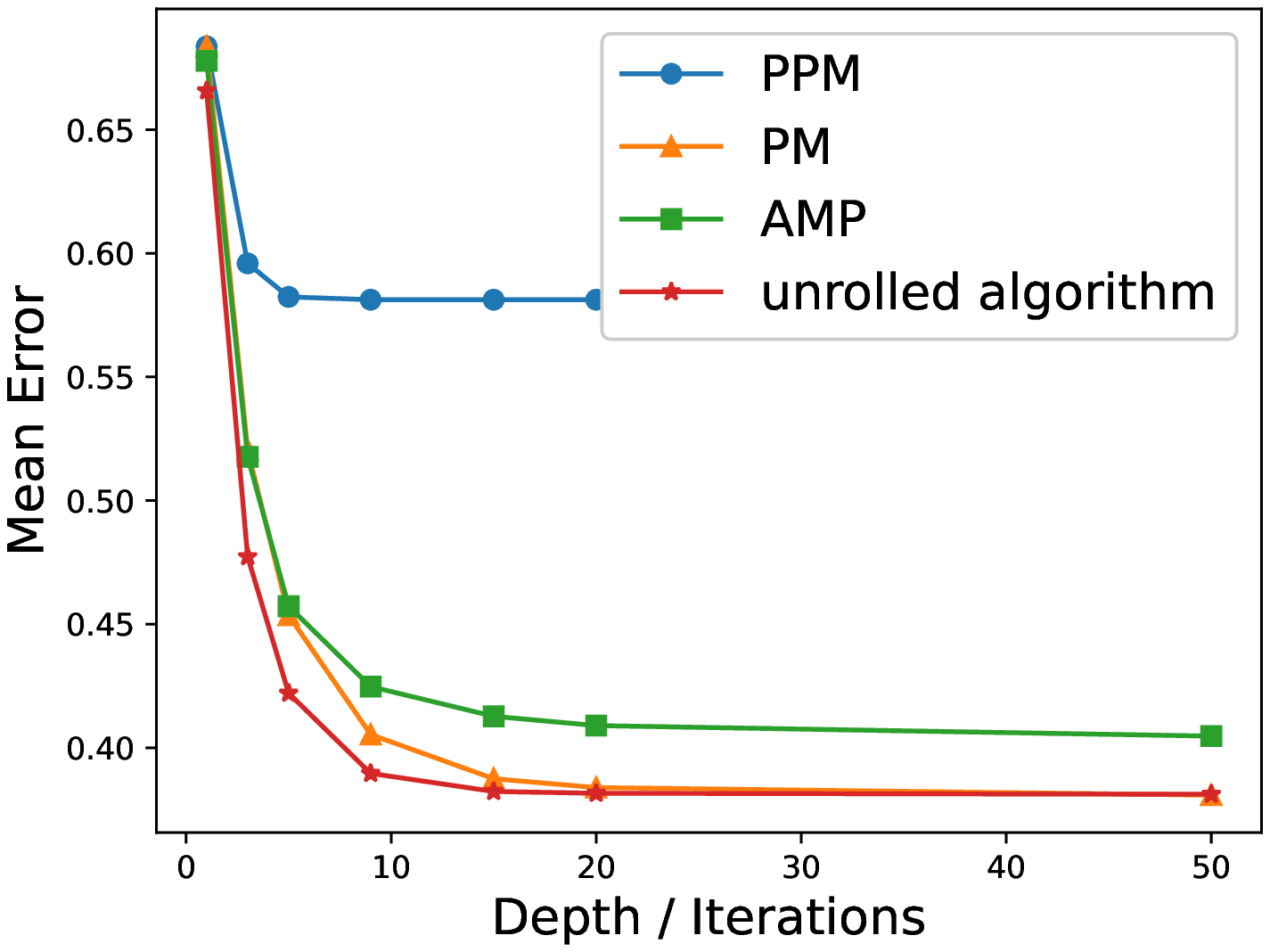}
		\caption{\(\lambda=0.3\)}
	\end{subfigure}
	\caption{\label{antiodal_alignment_loss}
Alignment error~\eqref{alignment_error_z_2} as a function of  depth for the multi-reference alignment over $\mathbb{Z}/2$  problem with different \(\lambda\) values. The unrolled algorithm is compared against the power method (PM), projected power method (PPM), and the AMP algorithm described in Section~\ref{mra_z_2_existing_methods}.
}
\end{figure*}

The network was trained using a dataset of \(M=10000\) samples, a batch size of $128$, with \(300\) epochs, and a learning rate of $10^{-4}$, using the Adam optimizer. 
The average alignment error as a function of depth is presented in Figure	\ref{antiodal_alignment_loss}, for \(\lambda=0.2\) and \(\lambda=0.3\). The experiment shows that the unrolled synchronization network usually 
achieves better error performance but the gap is insignificant. 

\subsubsection{With reconstruction loss \eqref{reconstruction_loss}}
\begin{figure*}
	\begin{subfigure}[ht]{\columnwidth}
		\centering
		\includegraphics[width=\columnwidth]{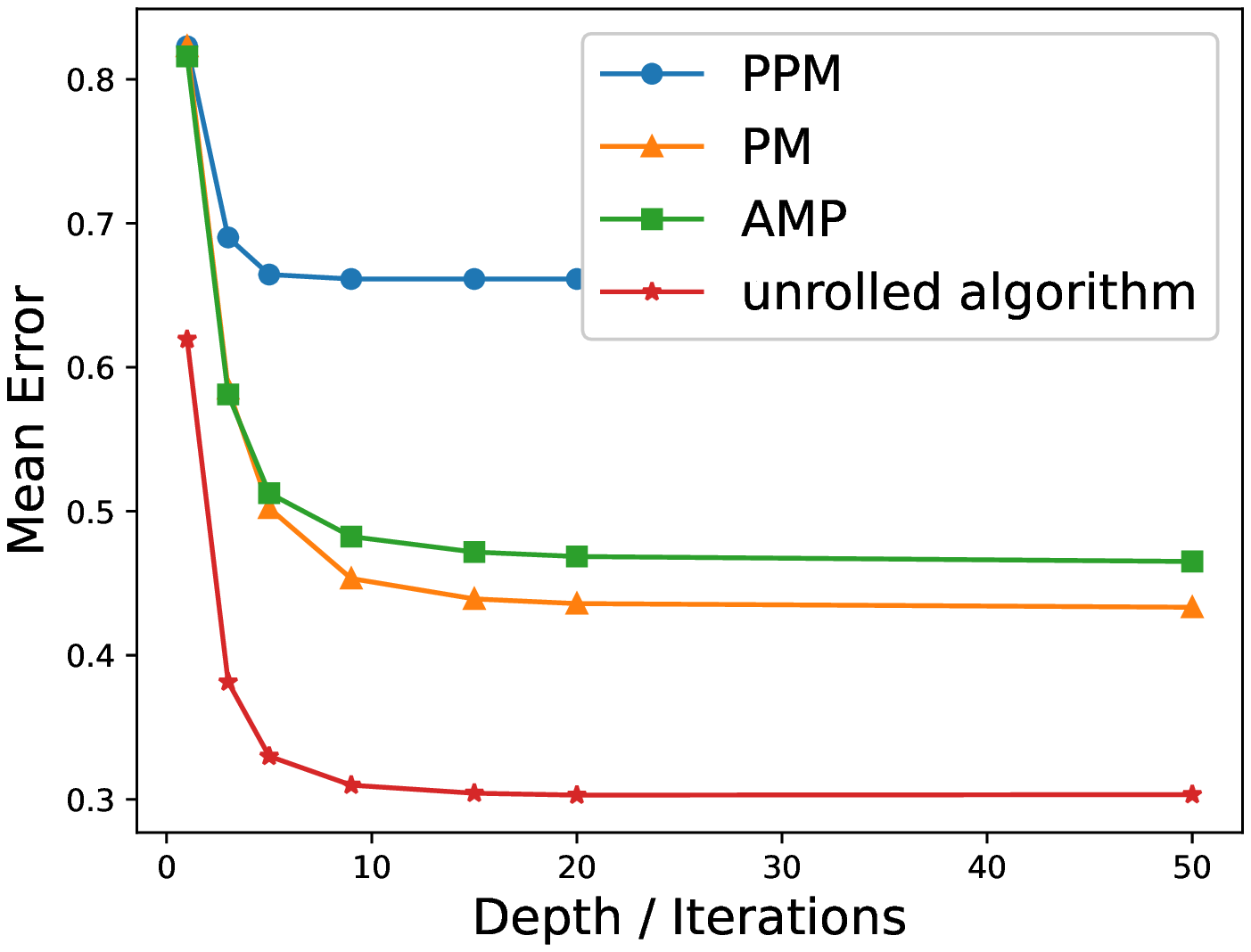}
		\caption{\(\lambda=0.4\)}
	\end{subfigure}
	\hfill
	\begin{subfigure}[ht]{\columnwidth}
		\centering
        \includegraphics[width=\columnwidth]{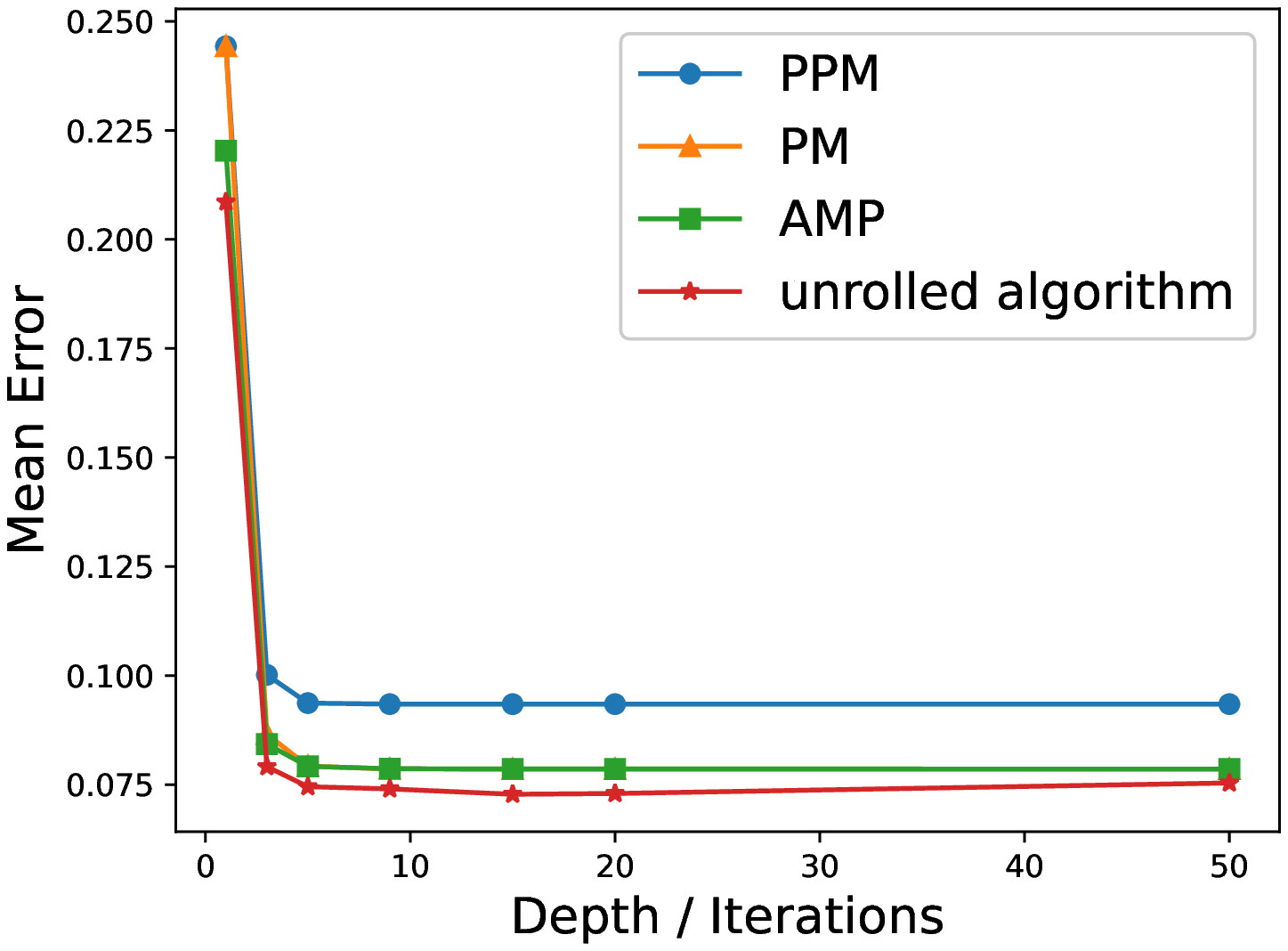}
		\caption{\(\lambda=0.8\)}
	\end{subfigure}
	\caption{\label{results_antipodal_rec_lambda_0p4}
Reconstruction error~\eqref{eq:rec_error_z2} as a function of  depth for the multi-reference alignment over $\mathbb{Z}/2$  problem with different~\(\lambda\) values. The unrolled algorithm, trained using the reconstruction loss function described in Section~\ref{sec:MRA_z_2_architecture}, is compared against the power method (PM), projected power method (PPM), and the AMP algorithm described in Section~\ref{mra_z_2_existing_methods}.
}
\end{figure*}

The network was trained using a dataset of \(M=10000\) samples, a batch size of $128$, with \(300\) epochs, and a learning rate of $10^{-3}$, using the Adam optimizer. 
The average reconstruction error as a function of depth is presented in Figure \ref{results_antipodal_rec_lambda_0p4} for \(\lambda=0.4\) and \(\lambda=0.8\). The experiment shows that the unrolled synchronization network achieves better reconstruction error performance per depth, and outperforms the existing methods for large number of iterations.

\subsection{Multi-reference alignment over the group $\mathbb{Z}/L$ of circular shifts }
We generated measurements according to~\eqref{1d_mra_model} with a signal of length  21, where each element was drawn i.i.d.\  
 from $\mathcal{N}(0,1)$, and \(N=20\).
The relative ratios were estimated according to \eqref{estimated_ratios_mra} and \eqref{ratios_matrix_mra}. In the first part, we evaluate the alignment error \eqref{alignment_error_u_1} using the network described in~\ref{sec:ArchitectureU1}. In the next part, we evaluate the signal reconstruction error, defined as:
\begin{equation}
\label{eq:rec_error_z_L}
	\text{error}(\mathcal{X},\hat{\mathcal{X}})=		\min_{\phi\in\{\frac{2\pi}{LP},2\frac{2\pi}{LP}...,2\pi\}}\|\mathcal{X} - e^{j\bar{k}\phi}\cdot\hat{\mathcal{X}}\|^2,
\end{equation}
where $\bar{k}$ is the frequency vector at each entry, $\cdot$ is an entrywise product, and  $\hat{\mathcal{X}}$ is the estimated signal in Fourier space, computed by aligning the measurements according to the estimated group elements and averaging, as described in \eqref{fourier_alignment} and \eqref{fourier_average}.
In this case, we used a modified loss function as described in \ref{sec:MRA_z_l_architecture}. We set \(P=10\).

\subsubsection{With alignment loss \eqref{alignment_loss_u_1}}
\begin{figure}[h]
	\includegraphics[width=\linewidth]{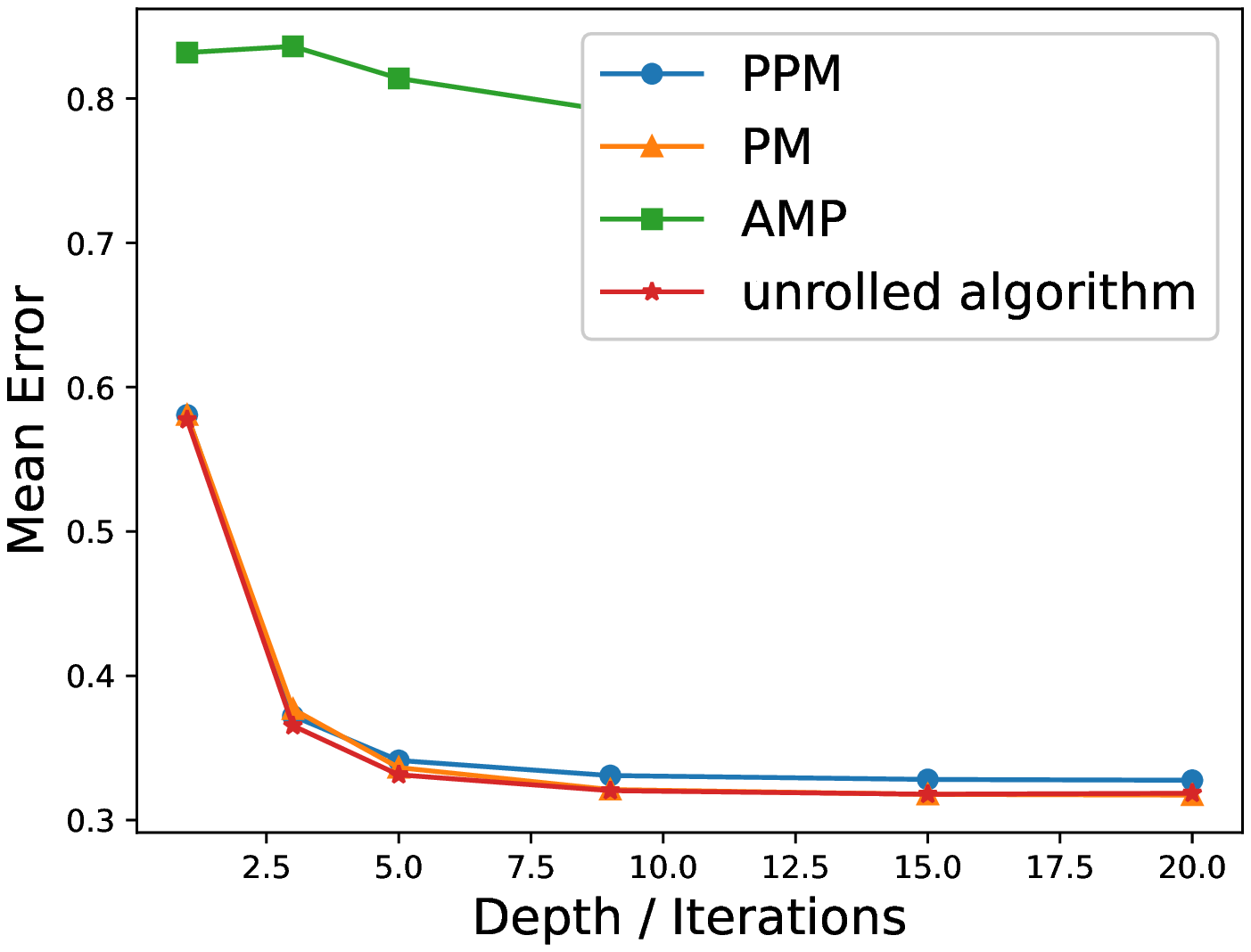}
	\caption{ 	\label{results_mra_1d_lambda_0p7}
Alignment error~\eqref{alignment_error_z_2} as a function of  depth for the multi-reference alignment over $\mathbb{Z}/L$ with \(\lambda=0.7\). The unrolled algorithm is compared against the power method (PM), projected power method (PPM), and the AMP algorithm described in Section~\ref{mra_z_l_existing_methods}.
	}
\end{figure}

The network was trained using a dataset of \(M=10000\) samples, a batch size of $128$, with \(300\) epochs, and a learning rate of $10^{-4}$, using the Adam optimizer. 
The average alignment error as a function of depth is presented in Figure~\ref{results_mra_1d_lambda_0p7} for \(\lambda=0.7\). 
The experiment shows that the error of the unrolled synchronization network improves with the depth of the network, but does not outperform the existing methods for large number of iterations.
\subsubsection{With reconstruction loss \eqref{reconstruction_loss_1d_mra}}
\begin{figure}[h]
	\includegraphics[width=\linewidth]{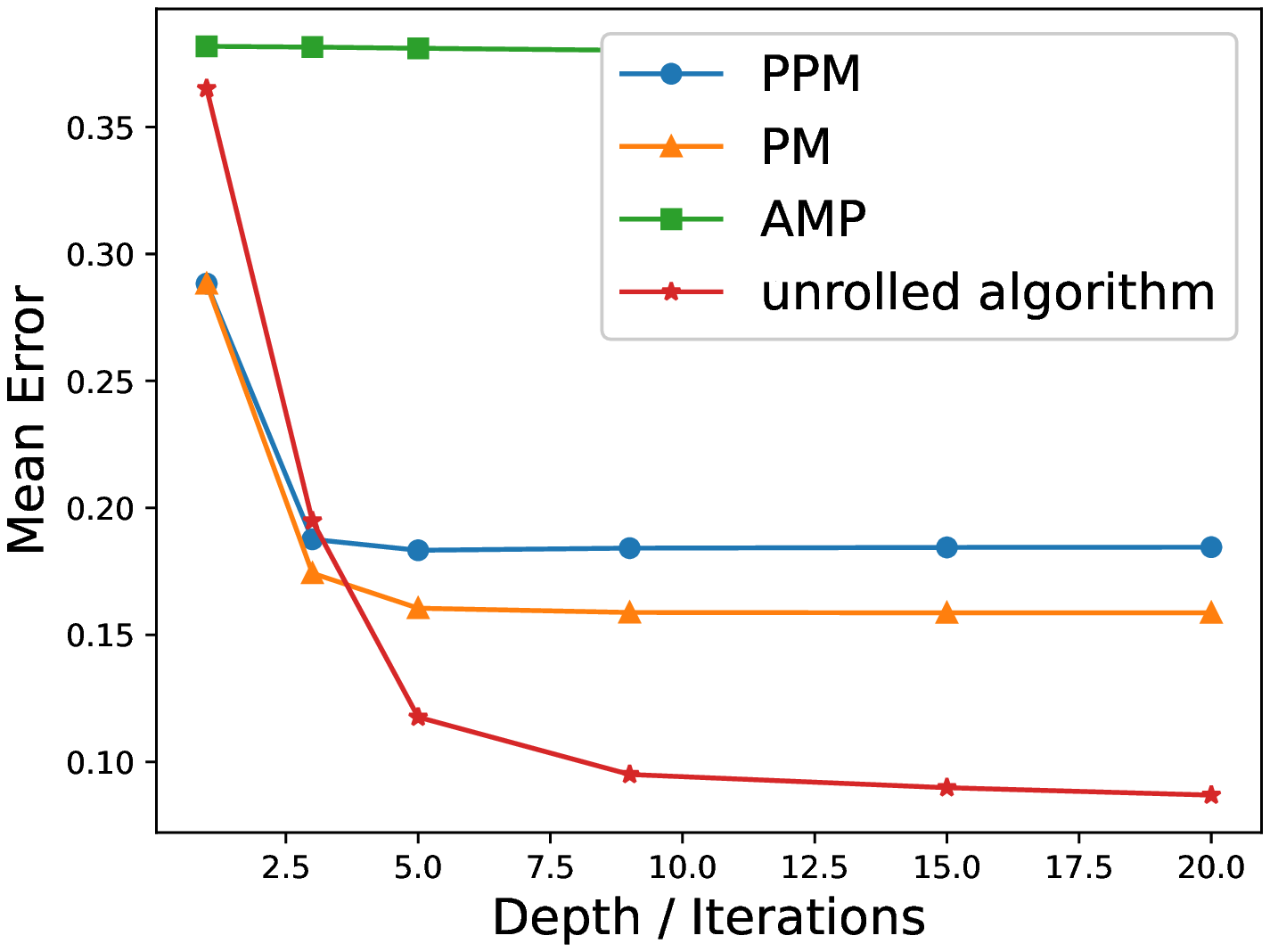}
	\caption{
Reconstruction error~\eqref{eq:rec_error_z_L} as a function of  depth for the multi-reference alignment over $\mathbb{Z}/L$  problem with~\(\lambda=1\). The unrolled algorithm, trained using the reconstruction loss function described in Section~\ref{sec:MRA_z_l_architecture}, is compared against the power method (PM), projected power method (PPM), and the AMP algorithm described in Section~\ref{mra_z_l_existing_methods}.
	}
	\label{results_mra_1d_lambda_1_reconstruction_loss}
\end{figure}

The network was trained using a dataset of \(M=10000\) samples, a batch size of $128$, with \(300\) epochs, and a learning rate of $10^{-1}$, using the Adam optimizer. 
The average alignment error as a function of depth is presented in Figure~\ref{results_mra_1d_lambda_1_reconstruction_loss} for \(\lambda=1\). 
The experiment shows that the unrolled synchronization network clearly outperforms the existing methods for large number of iterations.

\section{Discussion}
\label{sec:conclusion}

In this paper, we have presented a new computational framework for the group synchronization problem, based on unrolling existing synchronization algorithms, and optimize them using training data. 
\revv{We introduced unrolling strategies to a wide variety of group synchronization problems, trained using a differentiable invariant synchronization loss function that measures the alignment of the ground truth and the predicted group elements.
	We have shown that the designed algorithms outperform  existing methods for group synchronization. 	
For SO(3) synchronization, we suggested a differentiable feed-forward approximation for the projection operation which enables training  the unrolled algorithm.
For the MRA problem, the proposed algorithm incorporates  signal prior into the unrolling synchronization algorithm, since the training data consists of relative rotations estimated from noisy  measurements drawn according to the signal prior.
}

\rev{
In the $\mathbb{Z}/2$ synchronization problem, we have demonstrated how the suggested method achieves lower alignment error in the low and moderate SNR regimes, with fewer iterations.
 In the high SNR regime the performance of all algorithms is comparable, 
 but the unrolled algorithm still achieves a smaller error per iteration. 
 We then conclude that the proposed method is beneficial for lower SNR regimes,  and when running-time is a major concern. While existing methods such as AMP have asymptotic error guarantees, our experiments demonstrate that for a fixed and small number of samples the unrolled synchronization is favorable.
 We believe that the improved performance stems from our general strategy to optimize existing algorithms (such as AMP) using training data.
  Moreover, the unrolling synchronization can be readily applied to other noise models, beyond the Gaussian model.}

An interesting question is to examine whether a similar technique can be designed  for the non-unique games problem: a general optimization framework over groups that can be interpreted as a generalization of the group synchronization problem~\cite{bandeira2020non}.

The recent interest in the group synchronization and  MRA problems, and this paper in specific, is mainly motivated by the cryo-EM technology  to reconstruct 3-D molecular structures~\cite{bendory2020single}. 
In cryo-EM, each observation is a noisy tomographic projection of the molecular structure, taken from some unknown viewing direction. 
In particular, under some simplifying assumptions, the $i$-th cryo-EM observation is modeled as 
\begin{equation}
\label{cryoEM_model}
I_i = PR_{\omega_i}\phi +\varepsilon_i,
\end{equation}
where \(\phi:\mathbb{R}^3\to\mathbb{R}\) 
is the sought 3-D structure, \(R_{\omega}\) is a 3-D rotation by \(\omega\in\text{SO}(3)\), P is a fixed tomographic projection,  and \(\varepsilon\) is an additive noise. The goal is to estimate $\phi$, from $I_1,\ldots,I_N,$ while the rotations $R_1,\ldots,R_N\in SO(3)$ are unknown. 

One approach to solve the cryo-EM problem is to estimate the missing rotations from the observations
and then recover the 3-D structure as a linear problem. 
This methodology is  used to constitute \emph{ab initio} models~\cite{greenberg2017common}.
In \cite{vainshtein1986determination,van1987angular,singer2011three,shkolnisky2012viewing}, it was shown that  the pairwise relative rotations $\{R_{\omega_i}R_{\omega_j}^{-1}\}_{i,j=1}^N$ can be estimated from the observations based on the common lines property. Therefore, the cryo-EM reconstruction problem boils down to a synchronization problem over  SO(3).
Our ultimate goal is to apply our unrolled SO(3) algorithm to cryo-EM experimental data sets.
To train the network, in addition to simulated data as in this paper, we intend to use experimental data of previously resolved structures available in public repositories~\cite{iudin2016empiar}, 
 and structures resolved using computational tools such as AlphaFold~\cite{jumper2021highly}.

\revv{
Another possible future research thread is replacing the unrolling strategy  with deep equilibrium (DEQ) to enable a forward model corresponding to an infinite number of layers~\cite{DEQ}. Although  DEQ models were developed for sequence modeling, it may fit the group synchronization problem: the input sequence  is analog to the relative measurement matrix~$H$ that is shared among the layers, and the hidden sequence  is analog to the estimated group elements.
}
%
%
%
%

 \bibliographystyle{plain}



\end{document}